\documentclass[traditabstract]{aa}
\usepackage{longtable,lscape,amssymb}
\usepackage{graphicx}
\usepackage{tablefootnote}
\usepackage{amsmath} 
\usepackage{amssymb}
\usepackage{epsfig}
\usepackage{caption}
\usepackage{float}
\usepackage[caption=false]{subfig}
\usepackage{scrextend}
\usepackage{natbib}
\usepackage{array,booktabs}
\usepackage[colorlinks=true,citecolor=blue]{hyperref}
\usepackage{color}
\usepackage{ulem} 
\usepackage{adjustbox}
\usepackage[dvipsnames]{xcolor}

\newcolumntype{C}{>{$\displaystyle}c<{$}}

\bibpunct{(}{)}{;}{a}{}{,} 

\newcommand{\Ha}{{$H{\alpha}$}}
\newcommand{\Hb}{{$H\beta$}}
\newcommand{\Hg}{{$H\gamma$}}

\newcommand{\Hep}{{$H\epsilon$}}

\newcommand{\Msun}{{$M_{\odot}$}}
\newcommand{\Lsun}{{$L_{\odot}$}}
\newcommand{\Rsun}{{$R_{\odot}$}}
\newcommand{\vsini}{$v\sin i$}

\newcommand{\Mstar}{{$M_{\star} $}}

\newcommand{\Teff}{{$T_{\rm eff}$}}
\newcommand{\logg}{{$\log g$}}

\newcommand{\Lacc}{{$L_{\rm acc} $}}
\newcommand{\Macc}{{$\dot{M}_{\rm acc} $}}
\newcommand{\Ll}{{$L_{\rm line}$}}

\newcommand{\masszff}{2MASS J15551027-3455045}
\newcommand{\massffe}{2MASS J15414827-3501458}
\newcommand{\massfst}{2MASS J15361110-3444473}
\newcommand{\masstee}{2MASS J15523574-3344288}
\newcommand{\massztt}{2MASS J15383733-3422022}
\newcommand{\massttt}{2MASS J16011870-3437332}
\newcommand{\gaiants}{\textit{Gaia} DR2 6010590577947703936}
\newcommand{\gaiaefz}{\textit{Gaia} DR2 6014269268967059840}

\definecolor{dgreen}{rgb}{0.0, 0.27, 0.13}
\definecolor{blue-violet}{rgb}{0.54, 0.17, 0.89}

\begin{document}

\title{New members of the Lupus I cloud based on {\it Gaia} astrometry
\subtitle{Physical and accretion properties from X-Shooter spectra} \thanks{
Based on observations collected at the European Southern Observatory at Paranal, 
under program 105.20P9.001}
}

\author{
  F. Z. Majidi\inst{1,2}       
  \and J. M. Alcal\'a\inst{3}
  \and A. Frasca\inst{4}
  \and S. Desidera\inst{2} 
  \and C. F. Manara\inst{5}
  \and G. Beccari\inst{5} 
  \and V. D'Orazi\inst{2,6}
  \and A. Bayo\inst{5,7} 
  \and K. Biazzo\inst{8}
  \and R. Claudi\inst{2}
  \and E. Covino\inst{3}
  \and G. Mantovan\inst{1,2}
  \and M. Montalto\inst{4}
  \and D. Nardiello\inst{2,9}
  \and G. Piotto\inst{1}
  \and E. Rigliaco\inst{2}   }


\institute{        
  Dipartimento di Fisica e Astronomia, Universit\'a degli Studi di Padova, Vicolo dell’Osservatorio 3, 35122 Padova, Italy
  \and INAF-Osservatorio Astronomico di Padova, vicolo dell'Osservatorio 5, 35122 Padova, Italy \and INAF-Osservatorio Astronomico di Capodimonte, via Moiariello 16, 80131 Napoli, Italy
  \and INAF-Osservatorio Astrofisico di Catania, via S. Sofia, 78, 95123 Catania, Italy
 \and European Southern Observatory, Karl-Schwarzschild-Strasse 2, 85748 Garching bei M\"{u}nchen, Germany
 \and Department of Physics, University of Rome Tor Vergata, via della ricerca scientifica 1, 00133, Rome, Italy
 \and Instituto de F\'isica y Astronom\'ia, Facultad de Ciencias, Universidad de Valpara\'iso, Av. Gran Breta\~na 1111, Valpara\'iso, Chile
  \and INAF - Rome Astronomical Observatory, Via di Frascati, 33, I-00044, Monte Porzio Catone, Italy 
  \and Aix Marseille Univ, CNRS, CNES, LAM, Marseille, France}
\date{Received  }

\abstract{}


\abstract{We characterize twelve young stellar objects (YSOs) located in the Lupus I region, spatially overlapping with the Upper Centaurus Lupus (UCL) sub-stellar association. The aim of this study is to understand whether the Lupus I cloud has more members than what has been claimed so far in the literature and gain a deeper insight into the global properties of the region. We selected our targets using \textit{Gaia} DR2 catalog, based on their consistent kinematic properties with the Lupus I bona fide members. In our sample of twelve YSOs observed by X-Shooter, we identified ten Lupus I members. We could not determine the membership status of two of our targets, namely \textit{Gaia} DR2 6014269268967059840 and 2MASS J15361110-3444473 due to technical issues. We found out that four of our targets are accretors,  
among them \masszff, with a mass of $\sim$0.03\,\Msun, is one of the least massive accretors in the Lupus complex to date. Several of our targets (including accretors) are formed in-situ and off-cloud with respect to the main filaments of Lupus I, hence, our study may hint that there are diffused populations of M-dwarfs around Lupus I main filaments. In this context, we would like to emphasize that our kinematic analysis with \textit{Gaia} catalogs played a key role in identifying the new members of the Lupus I cloud.}

\titlerunning{New members of the Lupus I cloud}
\authorrunning{Majidi et al.}
\maketitle

\section{Introduction} 
\label{intro}

Observation of young stellar populations in nearby star-forming regions and 
comparison of their properties with more massive and distant ones is a key to understanding the impact of the environment on the star formation process and the properties of protoplanetary disks.

The Lupus dark cloud complex is one of the main low-mass star-forming regions (SFRs) within 200 pc of the Sun. It consists of a loosely connected group of dark clouds and low-mass pre-main sequence (PMS) stars. The complex hosts four active SFRs plus five other looser dark clouds with signs of moderate star-formation activity \citep{comeron2008}. Infrared (IR) and optical surveys \citep{evans2009, rygl2012}  have shown that objects in all evolutionary phases, from embedded Class I objects to evolved Class III stars, are found majorly concentrated in the Lupus I, II and III clouds with Lupus III being the richest in YSOs. Different distances to the Lupus stellar sub-groups have been claimed in the past from Hipparcos parallaxes and extinction star counts \citep{comeron2008}, but recent investigations based on \textit{Gaia} DR2 showed that the vast majority of YSOs in all Lupus clouds are at a distance of $\sim$160 pc \citep[see the Appendix in][]{alcala19}. Out of the three main clouds, Lupus III has been recognized as the most massive
and active star-forming region in Lupus by far, with a great number of young low-mass and very-low mass stars \citep{comeron2008}, while Lupus I, II and IV represent regions of low star-formation activity, with Lupus V and VI lacking star-formation \citep{spezzi11, manara18}. 

In this paper we investigate the Lupus I cloud. This cloud
has less than thirty bona fide members, which from now on we refer to as Lupus~I core members. The main motivation for selecting this cloud over the others with a low star-forming activity was the recent discovery of the star GQ Lup C \citep{alcala2020, lazzoni2020}, which is located on the main filament.
 
This target was specifically selected by our team for discovering possible
wide companions to SPHERE-GTO targets on \textit{Gaia} DR2 with a high specific interest in the presence of planets, brown dwarfs, or spatially
resolved circumstellar disks \citep{alcala2020, majidi2020}. GQ Lup C was proved to be a strong accretor that surprisingly had
escaped detection in previous IR and \Ha\ surveys, suggesting the possibility that many YSOs in the region are yet to be discovered. This discovery hence motivated us to conduct a more extended search in \textit{Gaia} DR2 to select new YSO candidates in the same region. In this work, we present the spectroscopic characterization of 12 YSOs in the Lupus I cloud.

The outline of this paper is as follows: in Sect. 2, we discuss the target selection criteria, as well as compiling a complete list of the bona fide Lupus I members, in addition to the observation and data reduction methods; in Sect. 3, we discuss the data analysis methods employed for analyzing the X-Shooter spectra, the membership criteria, and accreting objects; in Sect. 4, we discuss the results of our analysis; in Sect. 5, we introduce additional qualities of our targets in Lupus I, present their spectral energy distributions (SEDs), and evaluate them as potential wide companion candidates; and eventually, Sect. 6 will present our conclusions.

\section{Target selection, observations, and data reduction} 
\label{sec:obs}
\subsection{Target selection}
\label{sec:selection}
 
The \textit{Gaia} astrometric catalog \citep{gaia} has been recently used to efficiently identify young clusters and associations within 1.5\,kpc from the Sun \citep[see][and references therein]{prisinzano22}.
We selected our sample of YSO candidates based on a statistical analysis using the \textit{Gaia} DR2 catalog detailed in the following. As a first step, we identified the genuine population (core members) of Lupus I. These core members were gathered from the catalogs existing in the literature \citep{hughes1994, merin2008, mortier11, galli13, alcala14, frasca17, benedettini18, dzib18, comeron2013, galli2020}, and are listed in Table \ref{table:core}. We calculated the membership probability of these targets to Upper Centaurus Lupus (UCL)  with BANYAN $\Sigma$ \citep{gagne18} which are also quoted in Table \ref{table:core}. It should be noted that the catalog does not evaluate the Lupus membership. 

We then extracted the kinematic properties (i.e., parallaxes, $\varpi$, and proper motions $\mu_{\alpha*}$ and $\mu_\delta$) of these core members from \textit{Gaia} DR2, and constrained a range over these parameters \citep[see Appendix B of][]{alcala2020}. Using this constrained range, we searched for the objects with similar kinematic properties to Lupus I core members in \textit{Gaia} DR2 in a radius of 3 degrees from the center of the Lupus I cloud. At this stage, we found 247 objects. We placed these objects on a color-magnitude diagram (CMD) with Main Sequence (MS) stars \citep{pm13} and we removed those that were close to the limiting magnitude of Gaia (with photometric errors preventing a reliable classification according to their position on CMD) and we ended up with 186 targets. For generating this CMD, we used \textit{G} magnitudes and $Bp-Rp$ colors. This sample was then restricted to objects with a parallax within 5.5 to 7.5 mas (140-170 pc), within the $<\varpi>\pm4\cdot\sigma_{\varpi}$ parallax range of Lupus I core members, but we kept both sources lying close and far from the main filaments of the Lupus I to be inclusive both with the kinematic properties and spatial location of the selected targets. We also excluded those objects which were too faint for X-Shooter to observe ($J >$ 15 mag) or older than typical YSOs in Lupus I (inconsistent with the Lupus I core members on our generated CMD). 

\begin{figure}[ht!]
\includegraphics[width=\columnwidth]{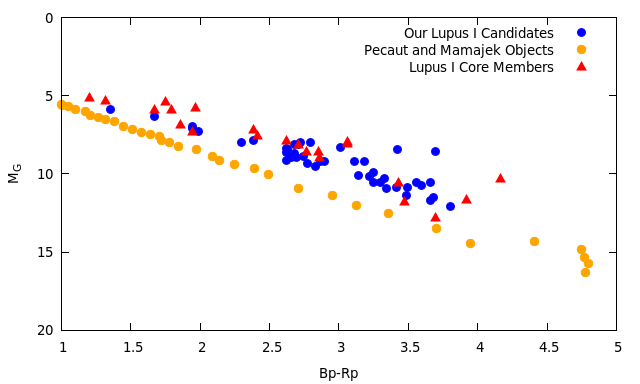}
    \caption{CMD of all the potential members of Lupus I in our original sample of 43 objects (blue dots), with the MS stars \citep[][]{pm13} (orange dots) and the Lupus I core members (red triangles) included in Table \ref{table:core}.}
    \label{fig:cmd}
\end{figure}

Taking into account all these constraints, we identified 43 candidates as potential members of Lupus I. As shown in the CMD in Fig. \ref{fig:cmd}, all of our eventual candidates lie above
the MS stars identified by \citet{pm13} and possess magnitudes and colors very similar to those of Lupus I members. Among these 43 objects, there are targets that i) have never been recognized as  potential members of Lupus I (17 objects), ii) were introduced as candidate members of Lupus\,I according to their consistent kinematic and/or photometric properties, but need spectroscopic confirmation (23 objects), iii) were known as members of Lupus\,I, but were poorly characterized in the literature, and, were never observed with X-Shooter (3 objects). We chose to include all these categories of objects to be followed up by X-Shooter, and the main reason for keeping the third category was that with X-Shooter spectroscopy we can determine their radial velocity (RV) and projected radial velocity (\vsini), or further explore their chromospheric and accretion properties in a more detailed fashion than previously done.

Targets in this category are Sz 70 \citep{hughes1994}, 2MASS J15383733-3422022 \citep{comeron2013}, and 
2MASS J15464664-3210006 \citep{eisner2007}. Among the eight objects selected in Lupus I in the unbiased photometric survey by \citet[][see their Table 2]{comeron2013}, only three were selected by our criteria and are those for which these authors provide stellar parameters, qualifying them as genuine YSOs. The other five were suspected to be foreground objects. Indeed, we confirmed that the astrometric parameters of the latter are out of range of our selection criteria.

As a final step, we cross-matched our full sample of 43 objects with the OmegaCAM \Ha\ survey in Lupus \citep[see][for details of this survey]{beccari18}, with only 4 being recognized as \Ha\ emitters.
This confirms that many potential YSOs may have escaped detection in \Ha\ imaging surveys and motivated us to spectroscopically characterize our full sample, giving a high priority to the four OmegaCAM \Ha\ emitters as potentially strong accretors.

\begin{table*}
        \centering
        \caption{Lupus I core members known from the literature (measurement errors are displayed in parenthesis). The column under Prob stands for the UCL membership probability percentage of the targets calculated by BANYAN $\Sigma$ \citep{gagne18}.}
        \begin{minipage}{\textwidth}
        \begin{tabular}{lccccccccr} 
        \hline\hline
        \smallskip
                Name & $\alpha$ (J2000) & $\delta$ (J2000) & $\varpi$ & $\mu_{\alpha*}$  & $\mu_\delta$ & RV & Prob & age \\
                & (h:m:s) & (d:m:s) & (mas) & (mas/yr) & (mas/yr) & (km/s) & \% & Myr \\
                \midrule
                
RX J1529.7-3628 &       																		       15 29 47.26 &	--36 28 37.41 &	6.04(0.09)	& --14.69(0.10) & --19.66(0.08) & 0.90(0.27)\footnote{\citet{gaia}\label{foot-gaia}} & 98.6 & - \\
IRAS 15334-3411	&	                           													   15 36 39.92 &	--34 21 42.17	& 6.89(0.13) &	--11.80(0.19) &  --19.84(0.12) & - &  91.6 & - \\

Sz 65/V$^{*}$ IK Lup	&                                          												   15 39 27.77 &	--34 46 17.21 &	6.44(0.05) &	--13.27(0.12) &	--22.24(0.07) & --2.70(2.00) & 98.6 & 1.9\footnote{Both RV and age are obtained by \citet{frasca17}\label{foot-frasca}} \\
Sz 66 &	                                           												   15 39 28.28 &	--34 46 18.09 &	6.36(0.09) & --13.60(0.19) &	--21.56(0.12) & 2.40(1.80) & 99.5 & 3.9\footref{foot-frasca}\\
RX J1539.7-3450A &            											   15 39 46.38 &	--34 51 02.54 &	6.40(0.04) &	--15.25(0.09) &	--22.33(0.05) & 7.17(1.28)\footref{foot-gaia} & 99.6 & - \\

UCAC4 274-081081 & 15 48 06.26 & --35 15 48.13 & 6.61(0.09) & --12.12(0.19) & --22.33(0.13) & - & 97.4 & - \\

RX J1539.7-3450B &																				   15 39 46.37 &	--34 51 03.66 &     6.40(0.13) & 	--13.52(0.26) & 	--20.85(0.13) & - & 98.2 & -\\

2MASS J15440096-3531056 & 15 44 00.96 & --35 31 05.72 & 6.45(0.14) & --11.49(0.26) & --24.07(0.19) & - & 89.3 & - 
\\
AKC2006 18 &	           													   15 41 40.81 &	--33 45 18.86 &	6.69(0.35) &	--18.84(0.33) &--22.06(0.27) & 9.10(2.30) & 95.3 & 8.3 \\
AKC2006 19 &	         							  15 44 57.89 &	--34 23 39.36 &	6.54(0.14) &	--18.94(0.089)& --22.75(0.06) & 9.60(2.10) & 97.0 & 8.0 \\

Sz 68/HT LUP A-B	& 																					   15 45 12.87 &	--34 17 30.65 &   6.49(0.06) & 	--13.63(0.13) & 	--21.60(0.08) & --4.3(1.8)&99.1&0.5\footref{foot-frasca}\\

HT Lup C &            																   15 45 12.67&	--34 17 29.37 &   6.55(0.19)& 	--15.43(0.22) & 	--20.27(0.15)& 1.2(3.9)\footref{RV}&97.8	& -\\

Sz 69	       &                                    												   15 45 17.41 &	--34 18 28.29 &	6.47(0.08) &	--15.05(0.15) &  --22.15(0.11) & 5.40(2.90) & 99.6 & 2.6\footref{foot-frasca} \\
2MASS J15451851-3421246 &	   												   15 45 18.52 &	--34 21 24.56 &	6.59(0.18) & --15.14(0.34) &  --21.77(0.22) & 4.40(2.90) & 99.7 & 0.5\footref{foot-frasca} \\
IRAS 15422-3414	 &                                  												   15 45 29.78 &	--34 23 38.81 &	6.46(0.17) &	--15.25(0.31) &	--22.52(0.24) & - & 99.1 & - \\

RX J1546.6-3618	&	                           													   15 46 41.20	& --36 18 47.44 & 6.69(0.07) &	--17.38(0.12) &	--24.29(0.08) & 7.20(0.10)\footnote{\citet{torres2006}} & 99.8 & - \\

Sz 71/GW LUP &	                                         											      15 46 44.73 &	--34 30 35.68 &	6.41(0.06) &	--14.03(0.10)	& --23.36(0.07) &  --3.30(1.90) & 99.0 & 2.0\footref{foot-frasca} \\

Sz 72/HM LUP	&                                         											  15 47 50.63 &	--35 28 35.40 &	6.41(0.05) &	--14.26(0.09) & --23.16(0.06) & 6.90(2.40) & 99.6 & 2.9\footref{foot-frasca}  \\
Sz 73/THA 15-5	                                         										 & 15 47 56.94 &	--35 14 34.79 &	6.38(0.06) &	--14.20(0.11) &  --22.26(0.07) & 5.00(2.20) & 99.7 & 3.7\footref{foot-frasca}  \\
GQ LUP/CD-3510525	                                         									 & 15 49 12.11	& --35 39 05.05 &	6.59(0.05) &	--14.26(0.09) &	--23.59(0.07) & --3.60(1.30) & 99.4 & 0.9\footref{foot-frasca}\\
Sz 76	         &                                  												   15 49 30.74 &	--35 49 51.42	& 6.27(0.05) &	--12.77(0.11) &  --23.37(0.08) & 1.40(1.00) & 99.4 & 2.3\footref{foot-frasca} \\
Sz 77	                           &                												   15 51 46.96	& --35 56 44.11	& 6.46(0.05) &	--12.42(0.09) &  --24.16(0.06) & 2.40(1.50) & 99.3 & 3.0\footref{foot-frasca}\\
RX J1556.0-3655	&	                           													   15 56 02.09 &	--36 55 28.27 &	6.33(0.04) &	--11.66(0.07) &	--22.50(0.05) & 2.60(1.20) & 99.3 & 7.8\footref{foot-frasca}\\               
2MASS J15443392-3352540\footnote{RV for this YSO candidate is the optimal RV determined by BANYAN $\Sigma$ as a member of UCL.\label{RV}} & 15 44 33.92 & --33 52 54.11 & 7.48(0.24) & --22.03(0.27) &--24.92(0.16) & 0.9(3.8)& 96.3 & 4.5\footnote{Age obtained by \citet{comeron2013}.\label{comeron}} \\   
2MASS J15392180-3400195\footref{RV} & 15 39 21.81& --34 00 19.56 & 6.39(0.19) & --17.23(0.2)& --20.18(0.15) & 1.1(3.8) &97.8& 7.1\footref{comeron}\\
\hline
             
 \end{tabular}
        \label{table:core}
\end{minipage}
\end{table*}

\begin{table*}[!h]
        \caption{Objects observed with X-Shooter (measurement errors are displayed in parenthesis). The column under Prob stands for the UCL membership probability percentage of the targets calculated by BANYAN $\Sigma$ \citep{gagne18}. The four candidates detected in the OmegaCAM \Ha\ imaging survey are flagged
        with ( \Ha ) right to their names (See Sect.~\ref{sec:selection}).}
        \begin{minipage}{\textwidth}
        \begin{tabular}{lccccccccr} 
                \midrule\midrule
                Name & $\alpha$ (J2000) & $\delta$ (J2000) & $\varpi$ & $\mu_{\alpha*}$  & $\mu_\delta$ & Prob & \textit{G} \\
                & (h:m:s) & (d:m:s) & (mas) & (mas/yr) & (mas/yr) & \% & (mag) \\
                \midrule \vspace{0.02cm}

\underline{Partially known targets:} & & & & & & \\ \\

2MASS J15383733-3422022 & 15 38 37.34 & --34 22 02.26 & 6.79(0.15) & --18.25(0.26) & --24.15(0.19) & 99.4 & 16.78  \\

Sz 70  & 15 46 42.99 & --34 30 11.55 & 6.09(0.21) & --12.58(0.39) & --22.16(0.25) & 95.7 & 14.50  \\

\hline \\
\underline{Candidates:} & & & & & & \\ \\

TYC 7335-550-1\footnote{Proposed candidate member of Lupus I by \citet{zari18}.} &  15 36 11.55 & --34 45 20.54 & 6.26(0.07) & --13.93(2.43) & --19.51(1.01) & 99.2 & 11.31 \\
2MASS J15361110-3444473\footnote{aka \textit{Gaia} DR1 6014141205925321984.} ( \Ha ) & 15 36 11.09 & --34 44 47.82 & 5.83(0.29) & --13.56(0.29) & --20.21(0.23) & 94.8 & 18.92 \\
2MASS J15523574-3344288\footnote{aka \textit{Gaia} DR2 6012155767105823616.} ( \Ha ) & 15 52 35.74 & --33 44 28.87 & 5.98(0.17) & --20.06(0.37) & --22.17(0.23) & 50.2 & 17.06   \\
2MASS J15551027-3455045\footnote{aka \textit{Gaia} DR2 6011827867821601792, candidate Lupus I member also proposed by \citet{galli2020}.} ( \Ha ) & 15 55 10.28 & --34 55 04.67 & 6.78(0.26) & --11.09(0.54) & --23.94(0.31) & 93.8 & 18.23  \\
2MASS J16011870-3437332\footnote{\textit{Gaia}  DR3 6011165313293141760.} ( \Ha ) & 16 01 18.70 &	--34 37 33.20 & 7.35(0.07) & --16.59(0.07) & --24.97(0.05) & 98.5 & 16.46 \\
UCAC4 269-083981\footnote{Dipper, candidate member of Lupus I also proposed by \citet{mimmo20}.} & 15 56 19.06 & --36 13 25.15 & 6.095(0.04) & --13.77(0.09) & --22.29(0.06) & 98.7	& 13.02 \\
\textit{Gaia} DR2 6010590577947703936 & 15 56 55.36 & --36 11 10.73 & 6.83(0.11) & --15.64(0.24) & --25.82(0.15) & 98.7 & 16.37  \\
2MASS J15414827-3501458\footnote{aka SSTc2dJ154148.3-350145, a candidate Lupus I member previously proposed by \citet{comeron2009}.} & 15 41 48.28 &	--35 01 45.84 &	6.74(0.13) & --17.99(0.25) &	--25.39(0.18) & 99.5 & 13.98 \\
UCAC4 273-083363 & 15 46 46.15 & --35 24 11.40 & 6.99(0.06) & --18.14(0.11) & --25.04(0.08) & 99.6 & 14.46  \\
\textit{Gaia} DR2 6014269268967059840 & 15 36 55.30 & --33 45 22.19 & 6.68(0.24) & --16.23(0.37) &	--22.29(0.27) & 95.3 & 17.39 \\

\midrule
        \end{tabular}
        \label{table:targets}
\end{minipage}
\end{table*}

\begingroup
\setlength{\tabcolsep}{2pt} 
\renewcommand{\arraystretch}{1.5} 
\begin{table*}[!h]
        \caption{Observing log of the new candidate members of Lupus~I. }
\begin{center}
\begin{scriptsize}
        \begin{tabular}{lcccccccccr} 
                \midrule\midrule
                Name & Date & Exposure time & Seeing & $T_{tot}$ & airmass & SNR & \textit{J} & Grade \\
                & (yyyy-mm-dd) & (sec) & (\arcsec) & (hour) & & & (mag)\\
                \midrule
                2MASS J15383733-3422022 & 2021-08-03 & 1920/1800/1920 & 1.72/1.72/1.72 & 0.67 & 1.04 & 5.4/47.1/68.6 & 13.39 & A\\
				Sz 70 & 2021-07-06 & 600/500/600 & 0.55/0.52/0.52 & 0.33 &  1.03 &  6.9/67.8/132.4 & 10.85 & A\\
				TYC7335-550-1 & 2021-06-27 & 300/200/300 & 0.72/0.77/0.77 & 0.33 & 1.36 & 71.1/117.0/245.6 &	  9.65 & A \\ 
				2MASS J15361110-3444473 & 2021-06-27 & 3600/3400/3840 & 0.73/0.69/0.70 & 1.25 & 1.15 & 0.1/4.9/21.3 & 14.91	& A \\ 
				2MASS J15523574-3344288 & 2021-06-27 & 1800/1700/1920 & 0.72/0.72/0.69 & 0.7 & 1.43 & 0.4/12.2/33.3 & 13.49 & A \\ 
         		
				2MASS J15551027-3455045 & 2021-08-01 & 1800/1700/1920 & 1.73/1.79/1.79 & 0.62 & 1.11 & 0.7/15.0/41.2 & 13.76 & A\\
         		
				2MASS J16011870-3437332 & 2021-08-08 & 1800/1700/1920 & 1.49/1.49/1.49 & 0.72 & 1.35 & 5.6/48.9/76.8  & 13.07 & A \\
         		
				UCAC4 269-083981 & 2021-08-01 &  600/500/600 & 2.27/2.27/2.27 &0.33 & 1.19 & 39.5/108.4/123.2 & 10.72 & C$^a$ \\
				
				\textit{Gaia} DR2 6010590577947703936 & 2021-08-06 & 1920/1820/1920 & 2.04/1.92/1.92 & 0.67 & 1.14  & 5.9/51.0/78.9 & 13.08 & A \\    
         		2MASS J15414827-3501458 & 2021-07-14 & 600/500/600 & 1.13/1.13/1.13 &  0.33 & 1.12 & 25.4/100.2/232.3 & 11.05 & A \\ 
         		UCAC4 273-083363 & 2021-07-14 & 600/500/600 & 1.33/1.29/1.33 & 0.33 & 1.08 & 18.3/73.6/171.0 & 11.55  & A\\
         		             				  			
           		 \textit{Gaia} DR2 6014269268967059840 & 2021-08-04 & 1800/1700/1800 & 2.49/2.49/2.49 & 0.65 & 1.13 & 1.5/26.1/50.5 & 13.64 & C$^b$\\ \midrule 
        \end{tabular}
\end{scriptsize}
\end{center}

\textbf{Notes.} Date of observation, exposure time allocated to each arm, mean seeing, and SNR (in order for UVB, VIS, and NIR wavelengths) as well as the total execution time, mean airmass, and the observation grades (as provided by the ESO observing staff) are reported.\\
$^a$ UCAC4 269-083981 had an out of constraint seeing (2\farcs0 which was exceeded). \\
$^b$ \textit{Gaia}~DR2 6014269268967059840 was reported to have an out of constraint seeing.\\
        \label{table:observation}
\end{table*}
\endgroup

\subsection{Observations}
\label{obs}

The observations were done with the X-Shooter spectrograph \citep[][]{vernet11} at the VLT, within a filler program, and terminated at the end of the observing period, when only $\sim$28\% of the
proposed sample was observed. Hence, of the 43 proposed targets, only 12 were eventually observed which are fully characterized in this paper, and are listed in Table \ref{table:targets}. The list of the targets that were not observed is reported in Appendix \ref{candidate-list}. 
These 12 targets were selected by ESO staff from the list of our proposed 43 targets, and include all of the \Ha \hspace{0.02 cm} emitters.
Although the observed sample is small, all the 12 observed targets were confirmed to be YSOs whose physical and chromospheric/accretion properties are worth to be investigated. For two stars the OBs were not validated by ESO observing staff (due to not fulfilling some of our requirements). But the spectra are nevertheless useful for classification purposes and are used in this work.

X-Shooter spectra are divided into three arms \citep{vernet11}, the UVB ($\lambda \sim$ 300--500 nm), VIS ($\lambda \sim$ 500-1050 nm), and NIR ($\lambda \sim$ 1000--2500 nm). We decided to observe all our targets with $1\farcs 0$, $0\farcs 9$, and $0\farcs 9$ slit widths (for UVB, VIS, and NIR arms respectively) for one or two cycles based on their \textit{J} band magnitudes. For our faintest objects with \textit{J} $>$ 14 mag, we considered two cycles of ABBA nodding mode. Among our observed targets, only 2MASS J15551027-3455045 belongs to this category, and due to its faintness, the final signal-to-noise ratio (SNR) of its spectra was lower than expected. The exposure time for each arm and the total execution time taking into account the overheads are reported for each target in Table \ref{table:observation}. For our brightest target, TYC7335-550-1 with \textit{J} = 9.65 mag, we decided that only one cycle of ABBA nodding would be sufficient for our scientific aims.

For some targets with a higher scientific significance to our program or because of their faintness, we decided to also observe telluric standard stars. Only a few of our targets (analyzed in this work) did not have a telluric star observation included in their observation block (OB) and these are UCAC4 273-083363, 2MASS J15414827-3501458 (with \textit{J} = 11.55 mag and 11.05 mag respectively), UCAC4 269-083981 (\textit{J} = 10.72 mag), and \textit{Gaia} DR2 6014269268967059840 (\textit{J} = 13.64 mag) which had a lower scientific priority for our program -- either were not lying on the main filament, were not strong candidates for membership in Lupus I, were not \Ha\ emitters, or were not faint for X-shooter to necessitate the observation of a telluric template. As we will detail later, we will also adopt a different approach to remove telluric lines for these objects. For the targets containing telluric observation in their OBs, the same nodding strategy as those of the targets was employed to minimize noise and cosmetics, with an airmass as close as possible to the targets. The airmass and seeing reported in Table \ref{table:observation} are averaged over the exposure times for each arm.

\subsection{Data reduction}    
The data used in this work have been reduced with the X-Shooter pipeline \textsc{xshoo} of version 2.3.12 and higher\footnote{\url{https://www.eso.org/sci/software/pipelines/xshooter/}}, and hence they have been de-biased, flat-fielded, wavelength-calibrated, order-merged, extracted, sky-subtracted and
eventually flux-calibrated. The result of this pipeline output is an ESO one-dimensional standard binary table and the two-dimensional ancillary files ready for scientific analysis. Flux calibration based on the photometric data available in the literature was done later directly on the available spectra, along with the telluric removal process which is not done for the distributed spectra reduced by the \textsc{xshoo} pipeline.

We used the Image Reduction and Analysis Facility \citep[IRAF,][]{iraf1, iraf2} to remove the telluric lines from the target spectra and to flux calibrate them, as well as to derive the stellar parameters from the spectra, which we shall discuss in detail in the upcoming sections.
Since the strategy for arranging our observation blocks did not include wide slit observations, the flux calibration of our targets totally relies on the photometric data available in the literature, which have been collected in various surveys (with the corresponding flux errors of e-16 $W.m^{-2}$ for the UVB arm, e-16 $W.m^{-2}$ for the VIS arm, and 2.5e-15 $W.m^{-2}$ for the NIR arm). For some of our faint objects, we only had access to very limited photometric data and had to calibrate the UVB portion of the spectra in accordance with the available photometric data in the VIS range. 

For the objects with observations of telluric standard stars, we removed the telluric lines and molecular bands using the IRAF task {\sc Telluric}. For the three targets without telluric star observations in our sample, which namely are 2MASS~J15414827-3501458, UCAC4 273-083363, and \textit{Gaia} DR2 6014269268967059840, we used the {\sc TelFit} Python code. This code fits the telluric absorption spectrum in the observed spectra \citep{gullikson14} using the LBLRTM code which models the line-by-line radiative transfer \citep{clough2005}. Applying {\sc TelFit}, we corrected the spectra for oxygen and water molecular bands in the visible range ($\sim$550-1000 nm), as well as for water, oxygen, and CO$_2$ molecular bands in the NIR ($\sim$1000-2500 nm) \citep[for the details on the wavelength ranges where these molecular bands dominate the spectrum the reader is referred to][]{smette15}.

\section{Data Analysis}
\label{sec:analysis-method}

There are several immediate aims that we planned to fulfill through our program. With the X-Shooter spectra, we can confirm the youth of the selected candidates through the presence 
of the \ion{Li}{i} (6708\,\AA) absorption line, in addition to \Ha\ emission, and other lines of the Balmer series as further hints. 
We also determine the spectral type (SpT) classification and the determination of stellar physical parameters such as effective temperature ($T_{\rm eff}$), luminosity ($L$), 
mass ($M$) and age. 
It is also possible that some of our candidates may belong to Scorpius-Centaurus Association (with an age 10-18 Myr, UCL sub-association)
rather than Lupus (1-2 Myr). We can single out these objects once we have fully characterized them.
The disentanglement between the two associations would be useful for clarifying their relationship. Using spectral lines of the Balmer series, we will also measure the accretion luminosity ($L_{\rm acc}$) and mass accretion rate ($\dot{M}_{\rm acc}$) of those objects that we qualify as accretors. In the following, we describe the methods used for achieving our immediate goals.

\subsection{Spectroscopic analysis methods}

\subsubsection{Spectral typing and line equivalent widths}
\label{subsec:spt}

To obtain the SpTs of our objects, we first compared the spectrum obtained with X-Shooter's VIS arm with a library of visible spectra of already characterized stars and brown dwarfs formerly observed by X-Shooter \citep[][]{manara13}. For the quantitative spectral typing of the stars, we then calculated the spectral indices described in \citet{riddick2007} based on the ratios of the average flux of molecular absorption bands within narrow wavelength regions, yielding in all cases an uncertainty of 0.5 subclasses. For TYC 7335-550-1 and UCAC4 269-083981, which are brighter than the rest of the targets and do not show clear molecular bands in their spectra suitable for measuring the Riddick's indices, the SpT is instead estimated through the $T_{\rm eff}$ obtained by the ROTFIT code (see Sect.~\ref{subsec:rotfit}). 
The results can be found in Table~\ref{stellarprop}.
 
The EW of the atomic lines reported in Table~\ref{table:EW} is measured by taking an average over i) the direct integration of the line profiles between two marked pixels and ii) fitting a Gaussian. The errors associated with these values thus report the difference between the measurements made with these methods. There are cases for which we could not detect the \ion{Li}{i} line at 6708 \AA. Hence, for these objects we only report an upper limit on the measurement of $EW_{\ion{Li}{i}}$. As suggested by \citet{cayrel1988}, a three-sigma upper limit on the flux of the lithium line can be calculated as:

\begin{table*}[!ht]
        \centering
        \caption{Physical stellar parameters of the targets obtained with the ROTFIT code. }
        \begin{minipage}{\textwidth}
        \centering
        \begin{tabular}{lcccccccccccr} 
                \midrule\midrule
                Name & $T_{\rm eff}$ &  log \textit{g} & vsin\textit{i} &  RV &  Prob  \\
                 & (K) & & (km/s) & (km/s) & \% \\
                \midrule
\massztt & 3111$\pm$70  &  4.75$\pm$0.13 &    $<$8 &  4.1$\pm$2.7 & 99.8	\\
	
Sz 70 & 3038$\pm$76  &  4.02$\pm$0.11  &  14.0$\pm$14.0  &   1.1$\pm$2.6 & 84.6	 \\
	TYC 7335-550-1 & 4488$\pm$140  &  4.06$\pm$0.22  &   $<$8  &     2.6$\pm$2.0 & 99.2 \\
	\massfst & 2883$\pm$104  &  4.41$\pm$0.12 &   13.0$\pm$10.0  &   6.9$\pm$2.6  		&97.9\\  
	\masstee & 2981$\pm$44  &  4.54$\pm$0.10     &    $<$8 &  2.6$\pm$2.7	& 75.3		\\   
	\masszff & 2700$\pm$103 &   3.60$\pm$0.11  &  19.0$\pm$8.0  &   0.1$\pm$2.9	& 97.9\\  
	\massttt & 3121$\pm$90  &  4.73$\pm$0.14  &  12.0$\pm$8.0  &  --0.5$\pm$2.3 & 98.7 \\
	UCAC4 269-083981 & 3846$\pm$47  &  4.53$\pm$0.11     &    $<$8 &      0.6$\pm$2.7	& 99.6		\\ 
	\textit{Gaia} DR2 6010590577947703936 & 3154$\pm$72 &   4.77$\pm$0.13 &   40.8$\pm$3.6    & 0.5$\pm$4.7	& 99.2	\\  
	\massffe & 3213$\pm$94 &   4.52$\pm$0.23 &   53.3$\pm$5.7  &   3.4$\pm$4.3		& 99.8\\
	UCAC4 273-083363 & 3211$\pm$56  &  4.51$\pm$0.15      &    $<$8 &   1.3$\pm$2.3 & 99.8	\\
	\textit{Gaia} DR2 6014269268967059840 & 3019$\pm$108  &  4.75$\pm$0.14 &   44.0$\pm$12.0 &    1.7$\pm$4.6 & 98.3 \\     \hline
     \noalign{\medskip}
     \end{tabular}
        \end{minipage}
        \label{stellarprop-rotfit}

\raggedright
    \textbf{Notes.} The column Prob represents the probability of the target to be member of Lupus~I according to BANYAN $\Sigma$, which is based on the RVs measured with ROTFIT and the kinematic properties reported by \textit{Gaia} DR2.
    \end{table*}

\begin{equation}
dEW = 3 \times 1.06 \sqrt{(FWHM) \hspace{0.02cm} dx} \hspace{0.02cm}/\hspace{0.02cm} (S/N) ,   
\end{equation}

\noindent in which FWHM is the full width at half maximum, S/N is the signal-to-noise ratio, and the bin size (dx) can be fixed to 0.2 \AA\ for the VIS arm. The values of these measurements are reported in Table \ref{table:EW} and Table \ref{table:EW-tyc} for TYC7335-550-1.   
  
  \begin{table*}[!ht]
        \centering
        \caption{EWs of the relevant lines indicating the chromospheric and accretion tracers for our targets. Negative values indicate the lines that are in emission.}
\begin{minipage}{\textwidth}
\begin{tiny}
\centering
        \begin{tabular}{lcccccccr} 
                \midrule
                Name & $EW_{\ion{Li}{i}}$  & $EW_{H{\alpha}}$ & $EW_{H{\beta}}$ & $EW_{H{\gamma}}$ & $EW_{H{\delta}}$ & $WH{\alpha}$(10\%)   \\
                & (\AA) & (\AA) & (\AA) & (\AA) & (\AA) & (km/s)  \\
                
                 \midrule
 
 \massztt & 0.74$\pm$0.04 & --8.77$\pm$0.92 & --7.71$\pm$0.04 & --7.99$\pm$0.21 & --7.20$\pm$0.52 & 128$\pm$18 \\
                
  Sz 70 & 0.55$\pm$0.05 & --43.37$\pm$3.97 & --9.97$\pm$1.07 & --10.28$\pm$1.04 & --11.14$\pm$1.51 & 366$\pm$14 \\      
\massfst & $<$ 0.25\footnote{\label{star}Three-sigma upper limits on the measurement (read Subsection \label{subsec:SpTy_EW} for further explanation).} & --71.4$\pm$8.77 & \dots & \dots & \dots & 292$\pm$14   \\ 
  \masstee &  0.81$\pm$0.09 & --13.52$\pm$0.76 &  --10.9$\pm$0.88 & --3.9$\pm$1.1 & --2.84$\pm$0.49 & 146$\pm$9 \\
 \masszff & -\footnote{Li I line was affected by a cosmic ray hit and could not be measured.} &  --88.9$\pm$1.17 & --29.7$\pm$0.85 & --6.68$\pm$0.24 & --5.09$\pm$0.49 & 229$\pm$14 \\
   \massttt & 0.67$\pm$0.03 &  --21.47$\pm$1.59 & --21.61$\pm$1.28 & --19.41$\pm$0.75 & --13.34$\pm$2.18 & 274$\pm$14 \\
  
  UCAC4 269-083981 & 0.56$\pm$0.01 & --1.69$\pm$0.07 & --1.63$\pm$0.08 & --1.56$\pm$0.24 &  --1.44$\pm$0.21 & 174$\pm$5  \\
 
   \gaiants &  0.68$\pm$0.06 & --6.53$\pm$0.38 & --6.75$\pm$0.25 & --6.97$\pm$0.09 & --6.69$\pm$0.22 & 183$\pm$5\\
   \massffe & $<$ 0.012\footref{star} & --10.04$\pm$0.53 & --9.55$\pm$0.61 & --10.64$\pm$0.29 & --10.21$\pm$0.7 & 210$\pm$18  \\
  UCAC4 273-083363 & $<$ 0.017\footref{star} & --11.4$\pm$0.94 & --11.12$\pm$0.45 &  --11.15$\pm$1.35 & --8.59$\pm$0.67 & 155$\pm$9 \\
  \gaiaefz & $<$ 0.047\footref{star} & --17.53$\pm$2.20 & \dots & \dots & \dots & 219$\pm$14	 \\
 \midrule
 \end{tabular}
        \label{table:EW}
        
\end{tiny}
\end{minipage}
\end{table*}

\begin{table*}[!ht]
        \centering
        \caption{EWs of the relevant lines indicating the chromospheric and accretion tracers for TYC 7335-550-1.}
\begin{minipage}{\textwidth}
\begin{tiny}
\centering
        \begin{tabular}{lcccccccr} 
 \midrule 
 Name & $EW_{\ion{Li}{i}}$ & $EW_{H{\alpha}}$ & $EW_{H{\epsilon}}$ & $EW_{\ion{Ca}{ii}}^{H}$ & $EW_{\ion{Ca}{ii}}^{K}$ &  $EW_{\ion{Ca}{ii}}^{8498}$ & $EW_{\ion{Ca}{ii}}^{8542}$ & $EW_{\ion{Ca}{ii}}^{8662}$    \\
                & (\AA) & (\AA) & (\AA) & (\AA) & (\AA) & (\AA)  & (\AA) & (\AA)  \\
                
 \midrule
 TYC 7335-550-1 & 0.39$\pm$0.02 & --0.45$\pm$0.06 & --0.32$\pm$0.16 & --1.07$\pm$0.14 & --1.41$\pm$0.19 & --0.47$\pm$0.03 &  --0.78$\pm$0.06 &  --0.68$\pm$0.06 \\
 \midrule
 \noalign{\smallskip}
 \end{tabular}
        \label{table:EW-tyc}
    ~\\
    \textbf{Notes.} The EW of \Ha, \Hep, and \ion{Ca}{ii} lines relate to the emission in the cores of these lines obtained by the subtraction of the photospheric template.   
\end{tiny}
\end{minipage}
\end{table*}

\subsubsection{ROTFIT}
\label{subsec:rotfit}

We used ROTFIT as the basis of our analysis for assessing the stellar parameters of our targets.
Using ROTFIT, we evaluated their RV, $v\sin i$, and surface gravity ($\log g$). The version of ROTFIT used for this purpose is the one designed for the optimal usage of the X-Shooter spectra \citep[][]{frasca17}. The stellar parameters obtained with ROTFIT can be found in Table \ref{stellarprop-rotfit}. 
The fitting process with ROTFIT code was carried out within a veiling (the UV excess continuum that influences the entire photosphere of the star from UVB to NIR) range from 0 to 1. None of our objects showed significant veiling, hence the veiling parameter for all our studied targets in this paper is equal to zero.

\subsubsection{Physical parameters}
\label{subsec:Phys_param}

\begin{figure}[!ht]
\centering
\includegraphics[width=9cm]{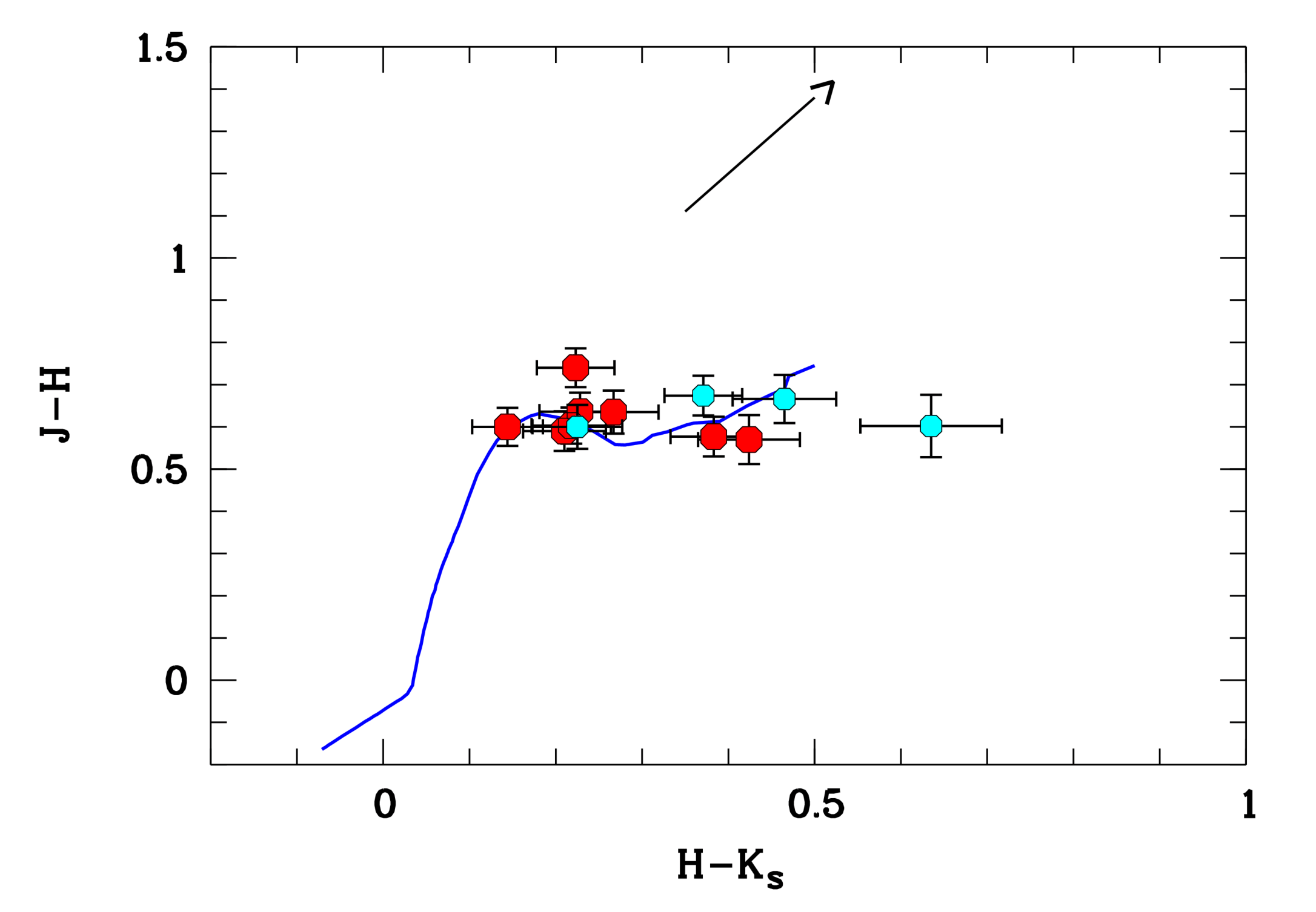}
\caption{ $J-H$ (mag) vs. $H-K_s$ (mag) diagram of all our targets. The red dots show the chromospherically-dominant targets, the cyan dots are the accretors, and the blue line represents the colors of MS objects, down to spectral type M9.5. The normal reddening vector, shown with the black arrow, corresponds to $A_V$ = 2 mag. The rightmost target is \massfst\ which is suspected to be a binary, hence, it might have color contribution from a second target.}
\label{fig:red-colors}
\end{figure}

We used the bolometric correction (BC) relation proposed by \citet{pm13, pm16} for evaluating the luminosity in both \textit{V} and \textit{J} bands and the radius of candidates according to their observed parallaxes and magnitudes. This is possible because none of our targets show significant near-IR excess (Fig. \ref{fig:red-colors}) nor strong veiling (Sect. \ref{subsec:rotfit}).

For the objects only resolved in \textit{Gaia} DR2 catalog, the BC relationship introduced by the \textit{Gaia} DR2 science team\footnote{\url{https://gea.esac.esa.int/archive/documentation/GDR2/Data_analysis/chap_cu8par/sec_cu8par_process/ssec_cu8par_process_flame.html}} is used. In order to have a correct estimation of the luminosity, we have also taken into account the extinction of the objects which was determined
using the grid of X-Shooter spectra of zero-extinction non-accreting T~Tauri stars \citep{manara13}, as explained in Sect.~3.2 of \citet{alcala14}. It is evident from Fig.~\ref{fig:red-colors} that the targets have low extinction and little or no NIR excess, probably except for the rightmost point in the diagram, which corresponds to \massfst. The relatively redder $H-K_{\rm s}$ color of this object in comparison with the others, may be due to the presence of an unresolved very late-type companion. This will be further discussed in Appendix \ref{app:notes-targets}.

\begin{table*}[!ht]
        \caption{Physical stellar parameters of the targets.}
        \begin{minipage}{\textwidth}
        \centering
        \begin{tabular}{lcccccccccccr} 
                \midrule\midrule
                Name & SpT 
                & $A_V$ & $L_{\star}$ & $R_{\star}$ & $M_{\star}$ &  Age & log \textit{g} \\
                 &  & (mag) & (\Lsun) & (\Rsun) & (\Msun) & (Myr)  \\
                \midrule
    \massztt & M5 
& 0 & 0.012$\pm$0.006&	0.39$\pm$0.01&	0.09$\pm$0.05&	10.7$\pm$5&	4.20$\pm$0.5 	\\
    Sz 70 & M5 
& 0.5 & 0.25$\pm$0.11 & 1.87$\pm$0.05 & 0.17$\pm$0.05 & 0.5$\pm$0.3 & 3.28$\pm$0.2 \\
	TYC 7335-550-1 & {\it K4.5} 
	& 0.7 & 0.94$\pm$0.56 & 1.60$\pm$0.05 & 1.1$\pm$0.1 & 3.50$\pm$1 & 4.04$\pm$0.2 \\
	\massfst & M5.5 
	& 1.75 & 0.006$\pm$0.003 & 0.32$\pm$0.01 &0.05$\pm$0.05	& 9.77$\pm$5 & 4.13$\pm$0.3		\\  
	\masstee & M5.5 
	& 0.5 & 0.02$\pm$0.01	&0.55$\pm$0.01 &	0.11$\pm$0.03 &	6.3$\pm$3 &	4.04$\pm$0.4			\\   
	\masszff & M7.5 
	& 0.75 & 0.0072$\pm$0.0034 &	0.39$\pm$0.02&	0.03$\pm$0.02&	1.7$\pm$1.5&	3.71$\pm$0.3			\\  
	\massttt & M5 
	& 0 & 0.013$\pm$0.006 &	0.41$\pm$0.01 &	0.09$\pm$0.04 &	9.55$\pm$5 & 4.16$\pm$0.5 \\
	UCAC4 269-083981& {\it M0} 
	& 0.5 & 0.30$\pm$0.14 &	1.23$\pm$0.02 &	0.6$\pm$0.3	&	4.2$\pm$1 &	4.03$\pm$0.5			\\ 
	\textit{Gaia} DR2 6010590577947703936 & M4.5 
	& 0 & 0.017$\pm$0.007&	0.45$\pm$0.01&	0.11$\pm$0.05&	8.8$\pm$4&	4.16$\pm$0.3		\\  
	\massffe & M4 
	& 0 & 	0.12$\pm$0.06 &	1.13$\pm$0.03&	0.2$\pm$0.08&	1.82$\pm$1&	3.64$\pm$0.4		\\
	UCAC4 273-083363 & M3.5 
	& 0	& 0.069$\pm$0.032&	0.83$\pm$0.01&	0.2$\pm$0.04&	3.63$\pm$1.5&	3.88$\pm$0.3				\\
	\textit{Gaia} DR2 6014269268967059840 & M6 
	& 0 & 0.01$\pm$0.005&	0.41$\pm$0.02&	0.05$\pm$0.03&	6.46$\pm$2&	3.93$\pm$0.5 \\     \hline
    \noalign{\medskip}
        \end{tabular}
        \end{minipage}
        \textbf{Notes.} The methods used for calculating SpT, $A_V$, $L_{\star}$, and $R_{\star}$ are described in the text. $M_{\star}$, log \textit{g}, and age of the stars are evaluated according to \citet{baraffe15} isochrones, except for TYC 7335-550-1, for which we have used the MIST isochrones. The SpT for TYC 7335-550-1 and UCAC4 269-083981 (in italic) are obtained using the temperatures derived by the ROTFIT code (Table~\ref{stellarprop-rotfit}) and the SpT--\Teff \hspace{0.02cm} calibration of \citet{pm13}. The errors associated with SpT and $A_{V}$ are 0.5 subclasses and 0.4 mag respectively. The errors associated with mass and age are internal to the tracks and isochrones.  
        \label{stellarprop}
\end{table*}

\begin{figure}[!ht]
\centering
\includegraphics[width=9cm]{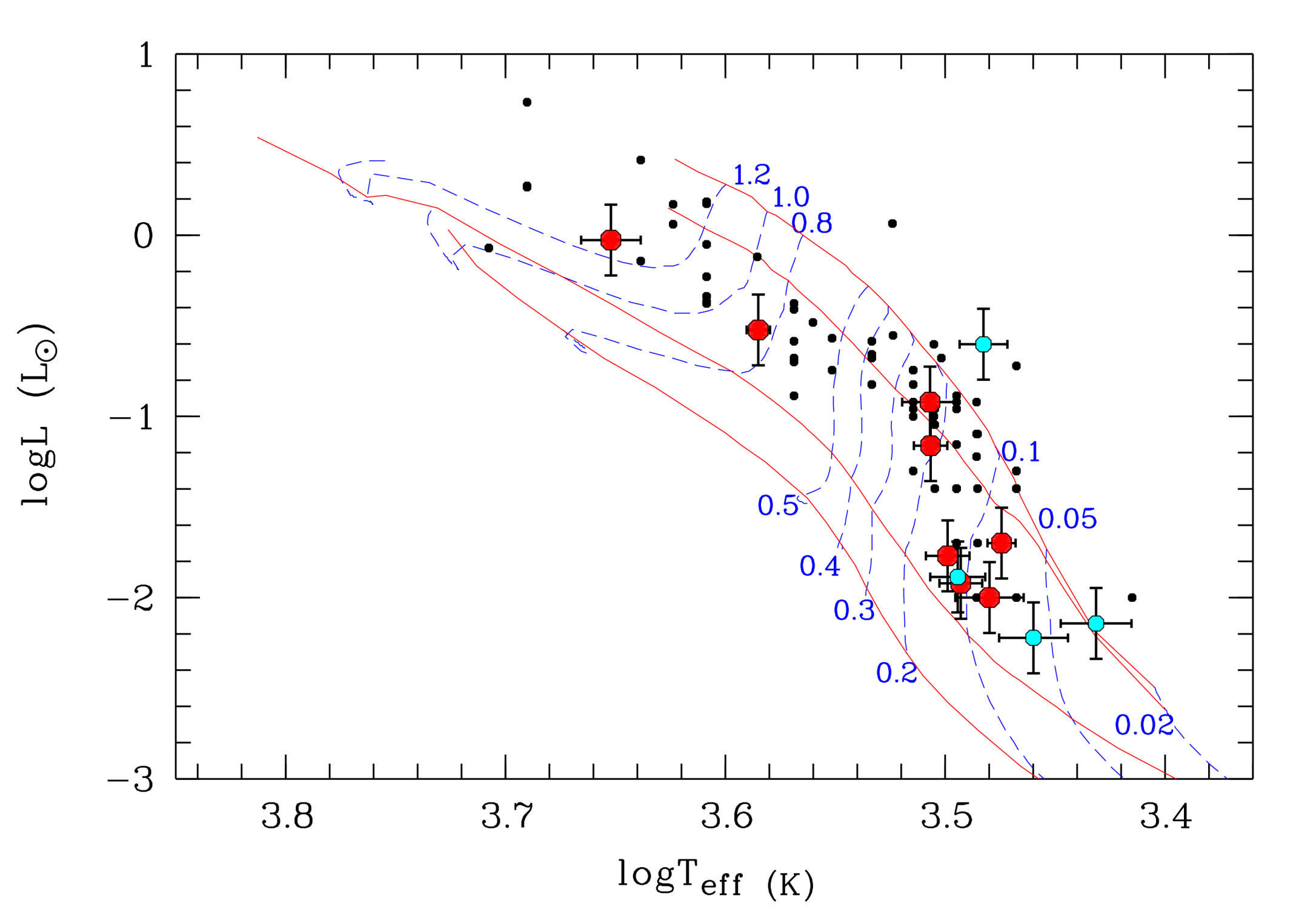}
\caption{$\log \hspace{0.02cm} L_\star (L_{\odot})$ vs $\log \hspace{0.02cm} T_{\rm eff}$ (K) diagram for all our targets (cyan and red dots represent accretors and non-accretors, respectively), together with the previously characterized Lupus members \citep[black dots,][sub-luminous objects are not plotted]{alcala19}. Blue dashed lines represent evolutionary tracks of \citet{baraffe15} for stars with masses indicated by the number (in $M_{\odot}$) next to the top or bottom of each track.
The red lines indicate
isochrones calculated with the same models at ages of 1, 3, 30 Myrs, and 10 Gyrs, from the right to the left.}
\label{fig:hrdiagram}
\end{figure}

Once the $T_{\rm eff}$ (from ROTFIT), luminosity, and radius of the targets are derived, their mass, age, and $\log g$ can be evaluated through various evolutionary tracks and isochrones available in the literature. The corresponding values of these parameters, which are reported in Table \ref{stellarprop}, are derived by the evolutionary models calculated by \citet{baraffe15}. The Hertzsprung-Russel (HR) diagram of the Lupus~I targets, including the previously known and the newly discovered members, is displayed Fig. \ref{fig:hrdiagram}. 
One of our targets, namely TYC\,7335-550-1, is much brighter than the other stars investigated in the present work, and falls outside the range covered by the \citet{baraffe15} models. Therefore, to derive its stellar parameters, we used MESA Isochrones and Stellar Tracks \citep[MIST][]{paxton15,choi16,dotter16}. For modeling purposes, we assumed that all targets have solar metallicity \citep{baratella20b}. 

Some of our objects display strong emission lines which is a sign of noticeable chromospheric activity (see the EW of some of the chromospheric activity indicators in Table \ref{table:EW}) or magnetospheric accretion from a circumstellar disk. If the magnetic activity is relevant, the position of the star in the HR diagram can be significantly affected by photospheric starspots and by the changes in the internal structure induced by the magnetic fields \citep[see][for interesting cases in the Taurus SFR]{gangi22}. In this case, isochrones that do not take into account these effects \citep[such as][]{baraffe15} may lead to systematic effects in the estimate of mass and age. In particular, they may indicate an age half the real age of star \citep{asensio19,feiden16}. This is crucial for our study which also aims at determining the membership of the stars in Lupus I or UCL associations. Thus, in addition to MIST and the isochrones provided by \citet{baraffe15}, we used other isochrones.  

A set of evolutionary models that considers the magnetic activity of the stars is the Dartmouth magnetic isochrones \citep{feiden16}, which we also use in this work to estimate the ages of all our targets. These isochrones were originally developed for estimating the age of the Upper Scorpius members (11$\pm$2 Myr), almost coeval to the UCL (15$\pm$3 Myr), and hence are quite useful to fulfill our scientific aims. 
In addition to \citet{baraffe15} and MIST models, we used both Dartmouth std and Dartmouth mag \citep[][and the references therein]{feiden16} models, as well as PARSEC + COLIBRI $S_{37}$ \citep{bressan12, pastorelli19, pastorelli20}. For all our targets, we obtained over-estimated ages using PARSEC + COLIBRI $S_{37}$ isochrones totally inconsistent with the other isochrones, hence, we do not report our results obtained with this isochrone to avoid confusion. The results of age estimation with all the other isochrones are included in Table \ref{table:age}. For all the models, we have assumed our targets have solar metallicity. For PARSEC models, extinction is also a free parameter that can be fixed and was thus set to the corresponding extinction of the targets reported in Table \ref{stellarprop}. Eventually, we would like to point out that it is not straightforward to state which targets may have an under-estimated age, particularly in the case of objects that are as young as the members of Lupus I and UCL considered in this work.

\subsection{Lupus I membership criteria}
\label{sec:membership-criteria}

According to the works previously done in the Lupus complex \citep[][and the references therein]{alcala14}, in addition to the kinematical properties expressed by the {\it Gaia} parallax and proper motions, membership criteria in this star-forming region are: 
 
i) the presence of lithium in their atmospheres, which is the main signature of youth. Despite the obviousness of this criterion, there are previously acknowledged members of the Lupus cloud that lack lithium. An example is represented by Sz 94 in the Lupus III cloud \citep{manara13,biazzo17,frasca17}; ii) an age 
consistent with the core members of the cloud. Although the estimated age of the Lupus complex is $\sim$ 1--2 Myr, there are previously recognized members of the complex that exceed this age range. Examples of such targets are AKC2006~18 and AKC2006~19 in Lupus I, although their apparent old age may be ascribed to disks seen edge-on that
obscure the central objects making them sub-luminous on the HR diagram \citep[see other examples in Sect.~7.4 in][]{alcala14}; iii) an RV consistent with the values of the
genuine members of the Lupus I \citep{frasca17}.

If an object does not match 
the membership criteria defined
above, there are two possibilities.
Either it is older than the UCL (age$>$20\,Myr), and we would hence identify it as field star; or it has a consistent age with UCL ($\sim$15\,Myr) which would confirm its membership to this sub-cloud of the Scorpius-Centaurus stellar association. To this aim, we have used various isochrones to evaluate the age of our targets.

\begin{figure}[!ht]
\centering
\includegraphics[width=9cm]{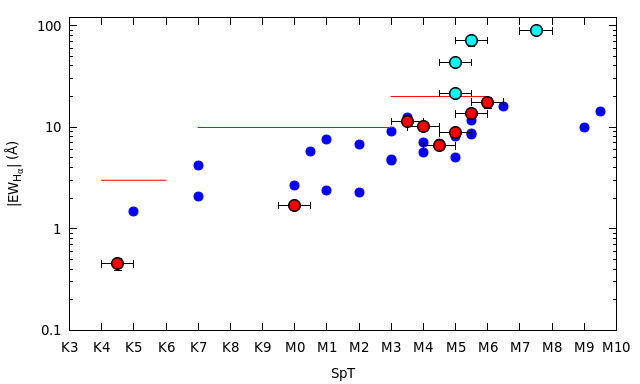}
\caption{$|EW_{H_{\alpha}}|$ vs SpT of our targets with the weak lined T Tauri stars studied by \citet[][blue dots]{manara13}. The cyan dots represent accretors, and the red dots represent chromospherically-dominant objects.
The horizontal lines in red represent the thresholds that separate non-accreting and accreting objects considering their SpTs \citep[][]{white2003}.}
\label{fig:ew-ha}
\end{figure} 
 
\subsection{Accreting objects}
\label{subsec:Accreting_obj}

There are several criteria for determining whether an object is actively accreting matter. Usually, 
an accreting object is characterized by
strong emission lines, strong UV and NIR continuum excess emission, or structured line profiles \citep[e.g.,][]{manara13}. Here, to establish whether an object is an accretor, we use the criterion proposed by \citet{white2003} which distinguishes the accreting and non-accreting objects based on the EW of their \Ha\ emission versus SpT. The method used in this paper for calculating the $L_{\rm acc}$ (accretion luminosity) and $\dot{M}_{\rm acc}$ (mass accretion rate) of our targets involves measuring the line luminosity of the emission lines of the accreting targets and using the established relationships between the $L_{\rm line}$ (for each emission line) with $L_{\rm acc}$ \citep[][]{alcala17}. We quote
the eventual accretion line luminosity that is obtained this way as $\log L_{\rm acc-line}$ in Table \ref{table:accretion-all} and Table \ref{table:accretion-tyc}.

The whole procedure that we carried out for this task can be summarized as follows: we corrected the spectra for telluric lines and flux-calibrated them, then measured the flux at Earth of the emission lines by integrating their profile above the local continuum, corrected the flux for extinction, calculated the luminosity of each emission line by multiplying the flux at Earth for $4\pi d$ (adopting a distance $d=1000/\varpi$\,pc, with $\varpi$ in mas), and eventually took an average over all the values of log $L_{\rm acc-line}$. We chose \Ha, \Hb, and \Hg\ emission lines to measure the accretion luminosity of our targets. After deducing the $\log L_{\rm acc}$ for each target, we obtained their $\dot{M}_{\rm acc}$ accordingly \citep{alcala17}. The results of our measurements are presented in Table \ref{table:accretion-all}.

\begin{figure}[!ht]
\hspace{-0.5cm}
\includegraphics[width=9.3cm]{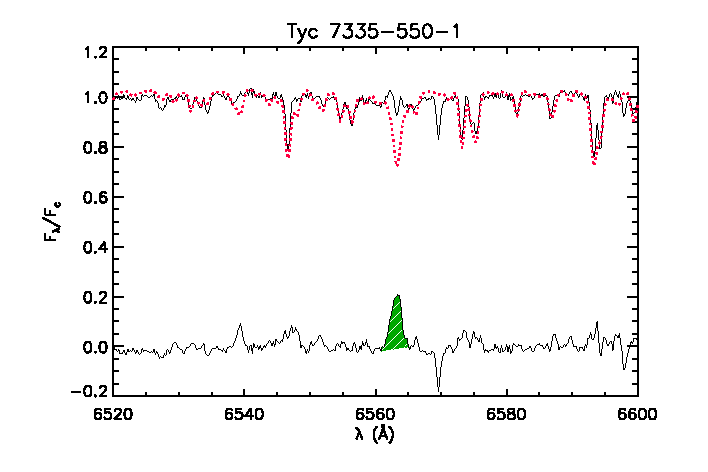}
\caption{X-Shooter spectrum of TYC\,7335-550-1 in the \Ha\ region, normalized to the local continuum (black solid line) along with the inactive photospheric template (red dotted line). The latter is produced by ROTFIT with the BT-Settl synthetic spectrum at the \Teff\ and \logg\ of this target that is degraded to the resolution of X-Shooter, rotationally broadened, and wavelength shifted according to the target RV.  The difference $target-template$ is displayed at the bottom of the box and emphasizes the \Ha\ emission that fills in the line core (green hatched area), which has been integrated to obtain the \Ha\ line flux. }
\label{fig:subHa}
\end{figure}

\begin{figure}[!ht]
\includegraphics[width=9cm]{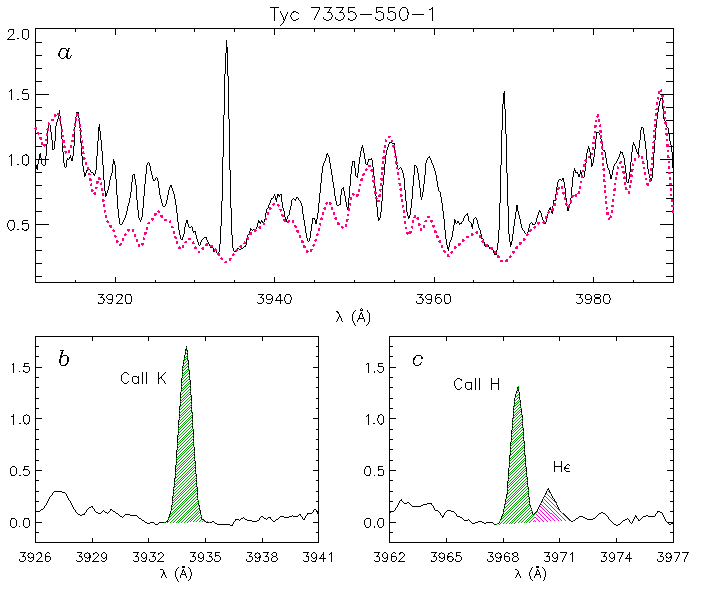}
\hspace{-.1cm}
\includegraphics[width=8.8cm]{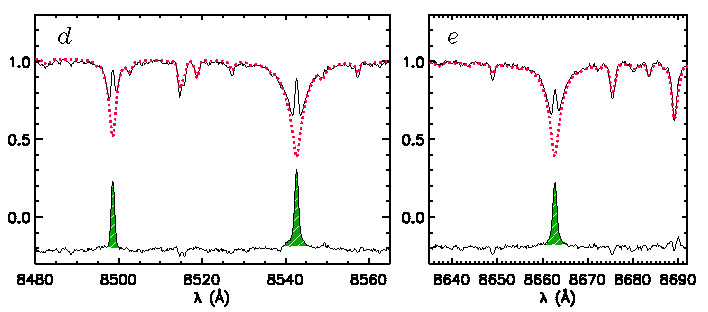}
\caption{{\bf a)} X-Shooter UVB spectrum of TYC\,7335-550-1 in the \ion{Ca}{ii} H\&K region (black solid line) along with the inactive photospheric template (red dotted line). {\bf b)} and {\bf c)} Residual ($target-template$) spectrum around the \ion{Ca}{ii}\,K and \ion{Ca}{ii}\,H line, respectively. The hatched green areas mark the residual H and K emissions that have been integrated to obtain the EWs and fluxes. The purple-filled area relates to \Hep. {\bf d)} and {\bf e)} Observed \ion{Ca}{ii} IRT line profiles (black solid lines) with the photospheric template overlaid with red dotted lines. The residual spectra are shown at the bottom of each panel shifted downward by 0.2 in relative flux units for clarity.   }
\label{fig:subCaII}
\end{figure}

Among all our targets, only TYC\,7335-550-1  does not show Hydrogen emission lines above the continuum, and its \Ha\ line is instead in absorption. For this target, we used ROTFIT to subtract the photospheric template in order to measure the flux of the emission components that fill the cores of Hydrogen and \ion{Ca}{ii} lines. This method has been successfully used to emphasize chromospheric emission or a moderate accretion whenever the photospheric flux is large and the emission is only detectable as a filling of the line core or an emission bump within the photospheric line wings that do not emerge above the continuum \citep[e.g.,][and references therein]{frasca15,frasca17}. 
The spectral subtraction allows us to recognize and measure the EW of the emission that fills in the \Ha\ line (Fig.~\ref{fig:subHa}). Adopting the same method, we measured the fluxes of the H\&K lines of the \ion{Ca}{ii} and in the cores of the three infrared lines of the \ion{Ca}{ii} IRT at $\lambda=$849.8, 854.2, and 866.2\,nm (Fig.~\ref{fig:subCaII}). We were also able to separate the contribution of the \Hep\ emission from the nearby \ion{Ca}{ii}\,H line.

\begin{table*}[!ht]
        \centering
        \caption{Accretion luminosity of the accretors derived from the line luminosities. The mass accretion rates are derived from the average of these values ($L_{acc-average}$).}
\begin{minipage}{\textwidth}
\begin{tiny}
\centering
        \begin{tabular}{lcccccccr} 
                \midrule\midrule
                Name &  log $L_{acc-H{\alpha}}$ & log $L_{acc-H{\beta}}$ & log $L_{acc-H{\gamma}}$ & log $L_{acc-average}$ & log $\dot{M}_{\rm acc}$     \\
                & (\Lsun)& (\Lsun) & (\Lsun) & (\Lsun) & (\Msun $yr^{-1}$)   \\
                
                 \midrule
\textbf{Accretors:}\\

  Sz 70 & --2.73 & --2.95 & --2.91 & --2.85 & --9.22   \\ 
  \massfst & --3.62 & \dots & \dots & --3.62 & --10.21  \\
  \masszff &  --3.85 & --3.95 & --3.96 & --3.92 & --10.20 \\
   \massttt & --4.04 & --4.29 &--4.20 & --4.16 & --10.91 \\
 \midrule 
 \textbf{Active stars:}\\
 \massztt & --5.41&	--5.43 &--5.52&--5.45&--12.21\\
\masstee & --4.62 &--4.87&--4.80&--4.75&-11.46\\
UCAC4 269-083981 & --4.07&--4.09&--4.24&	--4.13&-11.22\\
\gaiants & --5.12&	--5.09&--5.03 &--5.08 &--11.86\\
\massffe & --3.97&--3.93&--4.07&-3.99&-10.63\\
UCAC4 273-083363 &
--4.01&--4.14&--4.19&--4.11&--10.89\\
\gaiaefz & --5.22& \dots & \dots & --5.22& --11.07 \\

 \midrule
 \end{tabular}
        \label{table:accretion-all}
        
\end{tiny}
\end{minipage}
\end{table*}

\begingroup
\setlength{\tabcolsep}{2pt} 
\renewcommand{\arraystretch}{1.5} 
\begin{table*}[!h]
        \centering
        \caption{Accretion luminosity of TYC 7335-550-1 derived from its line luminosities. Its mass accretion rate is derived from the average of these values ($L_{acc-average}$).}
\begin{scriptsize}
        \begin{tabular}{lcccccccccr} 
        \midrule\midrule
                Name &  log 
                $L_{\rm acc}$& log 
                $L_{\rm acc}$ & log 
                $L_{\rm acc}$ & log 
                $L_{\rm acc}$ & log $L_{\rm acc}$ & log $L_{\rm acc}$ & log $L_{\rm acc}$
                & log $L_{\rm acc}$ & log $\dot{M}_{\rm acc}$     \\
                & \Ha & \Hep & Ca II (H) & Ca II (K) & Ca II (8498.02) & Ca II (8542.09) & Ca II (8662.14) & average\\
                & (\Lsun)& (\Lsun)& (\Lsun)& (\Lsun)& (\Lsun)& (\Lsun) & (\Lsun) & (\Lsun) & (\Msun $yr^{-1}$)   \\
                
                 \midrule
TYC 7335-550-1 & --3.43
& --2.82 & --2.31 &
 --2.19 &--2.01&--1.94 &--1.88 & --2.16 &--9.40
\\

 \midrule
 \end{tabular}
        \label{table:accretion-tyc}
\end{scriptsize}
\end{table*}
\endgroup

\section{Results}
\label{results}

\subsection{Stellar parameters and membership}
\label{subsec:param_memb}
The physical stellar parameters that we obtained from the spectral analysis and the HR diagram as described in Sects.~\ref{subsec:spt} and \ref{subsec:Phys_param} are reported in Table \ref{stellarprop}. 
The stellar parameters obtained with ROTFIT are presented in Table~\ref{stellarprop-rotfit}, where the membership probability 
was recalculated with the BANYAN $\Sigma$ using the values of RVs measured with ROTFIT. Both $T_{\rm eff}$ and log $g$ found with ROTFIT are in good agreement with those derived from SpT and the HR diagram and reported in Table \ref{stellarprop}. 

\begin{figure}[!ht]
\centering
\includegraphics[width=9cm]{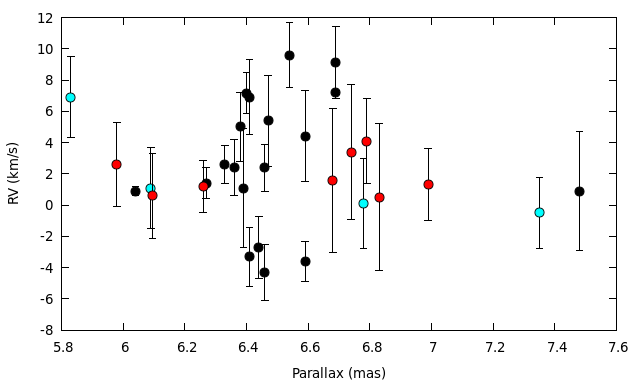}

\caption{RV of our accretors (cyan dots), chromospherically-dominant targets (red dots), and the Lupus I core members (black dots).
}
\label{fig:rv}
\end{figure}

We note that, at the resolution of the X-Shooter VIS spectra, the minimum value of $v\sin i$ that can be measured is 8~km/s (see, e.g., \citealt{frasca17}) 
and hence this value should be considered as an upper limit. With this knowledge, we can classify targets with $v\sin i <$ 8 km/s as slow rotators, and those with $v\sin i >$ 40 km/s as fast rotators. 
Moreover, the large RV range of the bona fide members of Lupus I ($\sim$ --5-12 km/s, according to Table \ref{table:core}) denies us to put a strict constraint on
the Lupus~I membership of our targets (Fig. \ref{fig:rv}). The RVs of the Lupus I members confirmed in this work, however, are within a smaller range with respect to the previously confirmed core members of the same region, except for \massfst\ which may or may not be a Lupus~I member. 

According to our full characterization, besides TYC~7335-550-1 which is a K4.5 type star, all the others have M spectral types. Three-quarters of our targets, have spectral types between M4 and M6, which is in accordance with the previously identified members of the Lupus complex 
\citep[]{alcala14, frasca17, krautter1997, hh14, comeron2013, galli2020}. The ages of these targets cover a large range of 0.7-11 Myrs, with masses in the range of 0.02 to 1.1 $M_{\odot}$ (as also indicated in Fig. \ref{fig:hrdiagram}).  

As discussed in Sect. \ref{sec:selection}, Sz 70 and \massztt\  were partially known in the literature. The physical parameters that we report here for Sz 70 are in excellent agreement with the results of \citet{hughes1994}. For \massztt, our results are again in good agreement with those reported by \citet{comeron2013}, but 
their difference emanates from the fact that \citet{comeron2013} measured $A_V$ = 1.2 mag for \massztt, which results in a discrepancy in luminosity, mass, and radius.

\subsection{Equivalent widths}
The EWs of several lines 
are quoted in Table~\ref{table:EW}, and separately for TYC~7335-550-1, in Table~\ref{table:EW-tyc}, as for this star the flux and EW measurements were performed by subtracting the photospheric spectrum.

\begin{table*}[!htb]
        \centering
        \caption{Overall status checklist for our targets. The rotation column refers to fast (F) or slow (S) rotators.}
        \resizebox{\textwidth}{!}{\begin{tabular}{lccccccccr} 
                        \midrule
                Name & Membership & Active & Accreting & Contains \ion{Li}{i} & Rotation & $A_{v}$ & Conclusion \\
                 & (UCL/Lup I) & (yes/no)& (yes/no)& (yes/no) & (F/S) & (mag) &\\
                \midrule
                2MASS J15383733-3422022 & Lup I & yes & no & yes & S & 0 & Genuine member of Lup I   \\

                Sz 70 & Lup I & yes & yes & yes & S & 0.5 & Genuine Lup I member +
                \\ & & & && &  &wide companion candidate  \\
TYC 7335-550-1 & Lup I & yes & no & yes & S & 0.7 & Genuine member of Lup I +     
\\ & & & && & &wide companion candidate  \\   
2MASS J15361110-3444473	& ? & yes & yes & no & S & 1.75 & Unresolved binary (?) + 		    
\\ & & & && & &wide companion candidate  \\
2MASS J15523574-3344288 & Lup I & yes & no & yes & S &  0.5 & New member of Lup I  \\
2MASS J15551027-3455045 & Lup I & yes & yes & ? & S & 0.75 & Genuine member of Lup I  \\
2MASS J16011870-3437332 & Lup I & yes & yes & yes& S & 0 & New member of Lup I  \\
UCAC4 269-083981 & Lup I & yes & no & yes & S & 0.5 &  Genuine member of Lup I \\
\textit{Gaia} DR2 6010590577947703936 & Lup I & yes & no & yes & F & 0 &  New member of Lup I \\
2MASS J15414827-3501458 & Lup I & yes & no & no & F & 0 & Genuine member of Lup I \\
UCAC4 273-083363 & Lup I & yes & no & no & S & 0 & Genuine member of Lup I\\
\textit{Gaia} DR2 6014269268967059840 & ? & yes & no & no & F & 0 & ? \\                      
                \midrule
        \end{tabular}}
\label{tab:status}
\end{table*}

We could not detect the \ion{Li}{i} line in the spectra of some of our targets for various reasons, which can be i) solely due to the low SNR of their spectra; ii) based on the simulations conducted by \citet{constantino21}, for initially lithium-rich stars we know that slow rotators could deplete their lithium (also considering their SpT) at early ages ($<$ 10 Myr), while fast rotators tend to retain their lithium; iii) a combination of the low SNR and fast rotation (which may be especially true for \textit{Gaia} DR2 6014269268967059840), which would further complicate the issues associated with \ion{Li}{i} detection; iv) a complex relationship between the accretion processes, early angular momentum evolution, and possibly planet formation for young stars ($\sim$ 5 Myr) that yet needs to be fully explored \citep{bouvier16}; v) no obvious relationship between the rotation of YSOs and the lithium depletion process \citep{binks22}.      

The non-detection of \ion{Li}{i} in the spectra of some objects has been reported as a three-sigma upper limit
on the flux of the lithium line which is a sensitive enough threshold for separating them from objects containing lithium.

\subsection{Evolutionary status of the targets}

The main properties and final status of all our targets are summarized in Table \ref{tab:status}. Based on all the criteria discussed in Sect. \ref{sec:membership-criteria}, we confirm that all our objects are YSOs, with ages $<$\,11 Myrs.

The targets 2MASS J15414827-3501458 and UCAC4~273-083363 do not show the presence of the lithium line in the spectra, but their effective temperature is compatible with the possible presence of a large amount of Li depletion for fully convective pre-main sequence stars (\citealt{bildstenetal1997}). Lithium depletion was investigated in several star forming regions, like some sub-groups of Orion (\citealt{pallaetal2007, saccoetal2007}), but also in Lupus I and III (see, e.g., \citealt{biazzo17}, and references therein). Due to their very young age ($< 4\,$Myr), we therefore classify 2MASS J15414827-3501458 and UCAC4~273-083363 as Lupus~I members. Newly discovered members of Lupus~I in this work are \masstee, \massttt, and \gaiants.  

There are also two objects analyzed in this work that we could not identify either as a member of Lupus I or UCL. These are 2MASS J15361110-3444473, whose spectrum indicates an unresolved binary star of spectral types M5.5 (VIS arm) and M8 (NIR arm), and we could not detect lithium in its spectrum (see Appendix \ref{app:notes-targets} for more details on the analysis of this target). However, we would like to emphasize that 2MASS J15361110-3444473 is an accreting source that has consistent kinematic and physical properties with the genuine members of Lupus I, hence, there is a possibility that this target also qualifies as a new member of Lupus I. The other object is \gaiaefz, for which we  
acquired a spectrum with poor SNR (see Sect. \ref{sec:obs} for details on the observation conditions of this target). The poor SNR of its UVB spectrum hindered us from carrying out any measurements on its \Hb\ and \Hg\ lines in emission (as reported in Table \ref{table:EW}), which also leads to evaluating its accretion properties only according to its \Ha\ emission line (as reported in Table \ref{table:accretion-all}). Therefore, the non-detection of lithium in its spectrum can be purely due the poor SNR in the VIS arm, and we do not approve nor rule out the possibility of this target being a member of Lupus~I.       

We hence confirm that all our targets are YSOs, with Hydrogen lines in emission above the continuum. Therefore, this investigation 
suggests that although only four of our targets were retrieved as \Ha\ emitters in the OmegaCAM survey (flagged in Table \ref{table:targets}), it is likely that
our entire sample of 43 candidate YSOs could include \Ha\ emitters or objects with filled \Ha\ profiles, which can only be confirmed by a high- or mid-resolution spectroscopic study or in deep X-ray surveys.

As a further investigation to strengthen our argument, we cross-matched all of the Lupus I core members included in Table \ref{table:core} with the OmegaCAM survey. Except for three objects, they were all retrieved in the survey as \Ha\ emitters. These exceptional three core members are RXJ1529.7-3628 (which was out of the field of view of the survey), RX J1539.7-3450B and Sz~68/HT~Lup~C, for which only one object was resolved in the survey. Combining this result with the results of this paper, we emphasize the necessity of observing all our sample to characterize all the members of Lupus I that have escaped the \Ha\ surveys. 

\subsection{Accretion versus chromospheric--dominated objects}
\label{sec:accretors}

\begin{figure}[!ht]
\centering
\includegraphics[width=9cm]{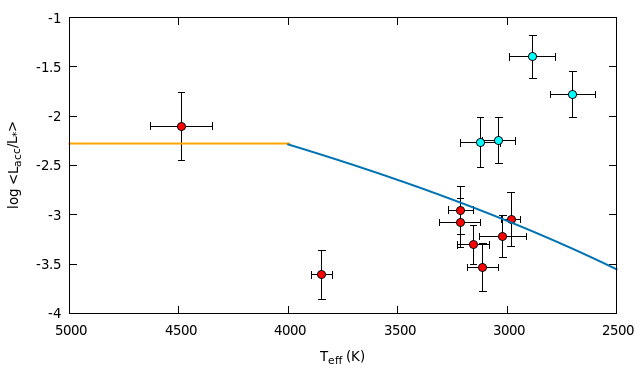}
\caption{Log $<L_{acc}/L_*>$ vs $T_{\rm eff}$ for all our targets. The cyan dots represent accretors, and the red dots represent chromospherically-dominant targets. The lines indicate the limit below which the chromospheric activity for a star is dominant \citep{manara17b}, for two regimes of stars with $T_{\rm eff} \leq 4000$ K (the diagonal blue line) and those with $T_{\rm eff} \geq 4000$ K (the horizontal orange line).}
\label{fig:accretors}
\end{figure}

We realized that four of our targets in the current sample are accretors. We measured the $L_{acc}$ of these targets, in addition to our chromospherically-dominant objects (Table \ref{table:accretion-all} and Table \ref{table:accretion-tyc}). The measured $L_{acc}$ for all our targets are displayed in Fig. \ref{fig:accretors}.
In the same figure, we have included the limits suggested by \citet{manara17b} for objects with $T_{\rm eff} >$ 4000 K and $T_{\rm eff} <$ 4000 K, below which the chromospheric activity of targets is dominant. All our four accretors exceed this limit for targets with $T_{\rm eff} <$ 4000 K, confirming that they are accretion-dominated. The rest of our targets within the same effective temperature range are below this threshold, which make them chromospheric-dominated objects, as expected. \masstee, however, lies exactly on the threshold between these two regimes, which is consistent with its significant \Ha\ emission. We also emphasize that this target was retrieved in the OmegaCAM survey as an \Ha\ emitter. 

\begin{figure}[!ht]
\centering
\includegraphics[width=9cm]{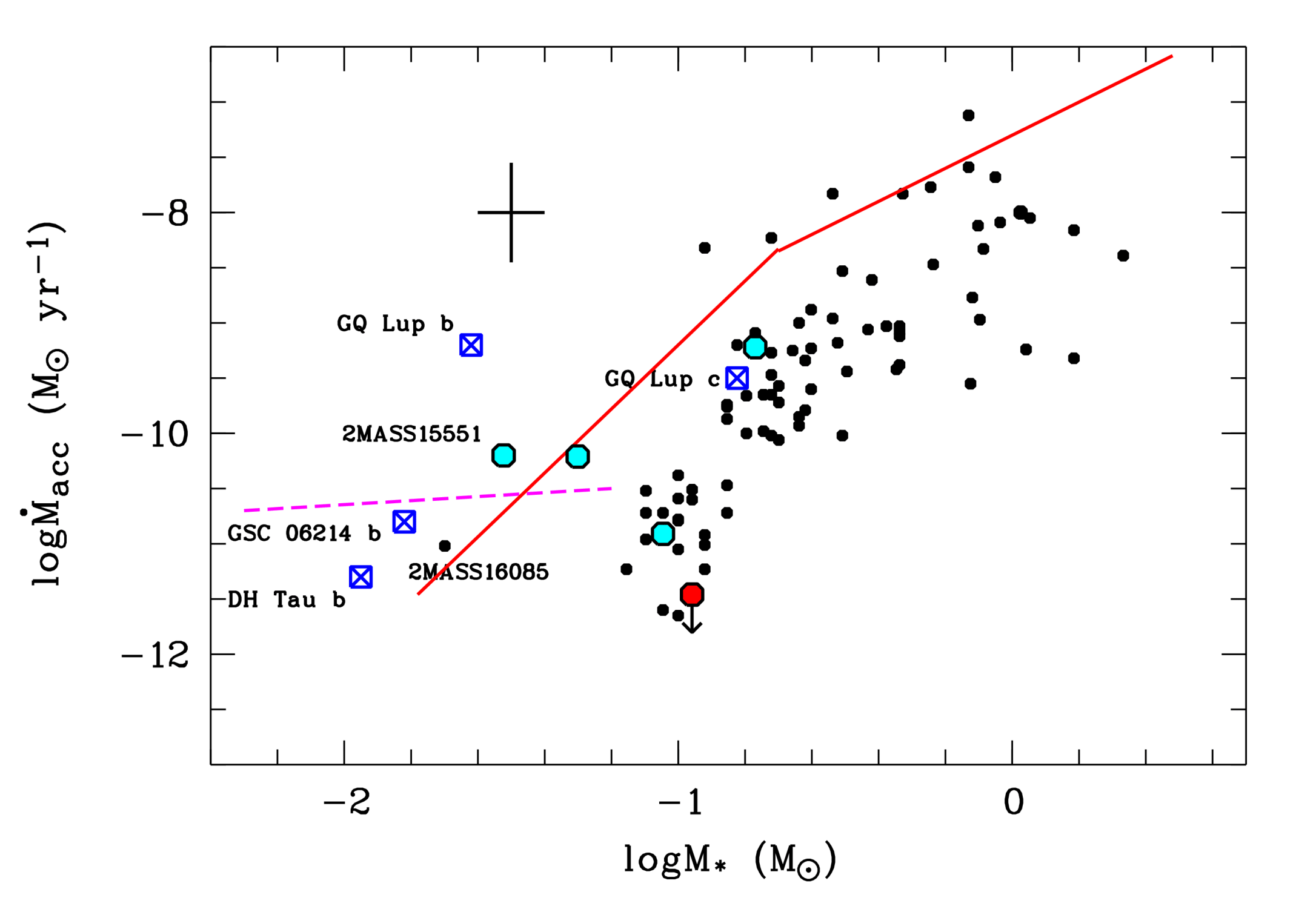}
\caption{Log $M_{acc} (M_{\odot}/yr)$ vs log $M_* (M_\odot)$ for the four accretors in our sample (cyan dots), together with the previously identified members of the Lupus (black dots). The blue crossed squares represent the substellar accreting companions detected at wide orbits by \citet{zhou2014} around GQ\,Tau, GSC\,06214\,00210 and DH\,Tau as labeled. \masszff, GQ\,Lup~c and 2MASS\,J16085953-3856275 are also labelled. \masstee \hspace{0.02cm} is labelled as red dot.
The continuous red line indicates the double power-law prediction of \citet{vb2009}, while the magenta dashed line shows the prediction of disk
fragmentation model by \citet{staher15}.}
\label{fig:mass-accretion}
\end{figure}

Fig. \ref{fig:mass-accretion} shows the $\dot{M}_{\rm acc}$ versus $M_*$ for the four accretors in our sample in comparison with the Lupus members. 
Among the four accretors, \masszff  ~is the least massive target, and has a very high mass accretion rate in comparison with Lupus members of similar mass. This target also stands above the double power-law relationship between $\dot{M}_{\rm acc}$ and $M_*$ established by \citet{vb2009}, based on modeling self-regulated accretion by gravitational torques in self-gravitating disks. As concluded by \citet{alcala17}, only the strongest accretors stand above this model. Our three other accretors have values of mass accretion rates typical of Lupus
accretors.

Finally, it is worth noting that three of our accretors (Sz 70, \massfst, and \massttt) have $WH_{\alpha}$(10\%)$>$270 km/s (see Table~\ref{table:EW}), which is expected from accreting stars. Our chromospherically-dominant targets have much narrower \Ha\ profiles.

\section{Discussion}
\label{diskussion}

In this paper, we analyzed 12 objects observed by X-Shooter out of our original sample of 43 proposed new candidate members of Lupus I. We confirm that all these 12 objects are YSOs, and ten out of 12 are members of Lupus I. We could not determine the membership of two of our targets, namely 2MASS J15361110-3444473 and \gaiaefz, as explained in the previous Section. We could not fully measure the accretion properties of \gaiaefz\, and hence our analysis in this regard for this specific target is not reliable. 2MASS J15361110-3444473, on the other hand, is a rather (intrinsic) faint object to be followed up by any available spectrographs, but perhaps can be followed up with ALMA to understand whether it is surrounded by a disk. Although recognized to have an older age with respect to Lupus I members (9 Myr), it can be still strongly accreting matter, consistent with the members of $\gamma$ Vel with age $\sim$10 Myr \citep{frasca15}. One of the interesting targets discussed in this work is TYC 7335-550-1, a lithium-rich K-type star with \Ha\ in absorption and without IR excess. We would like to emphasize that YSOs with these particular characteristics would never appear in \Ha\ imaging surveys such as OmegaCAM, although one of their main aims is to identify the members of young star forming regions. All the above points considered, we have fully characterized ten members of Lupus I in this work.

\begin{figure*}[!ht]
\includegraphics[width=9cm]{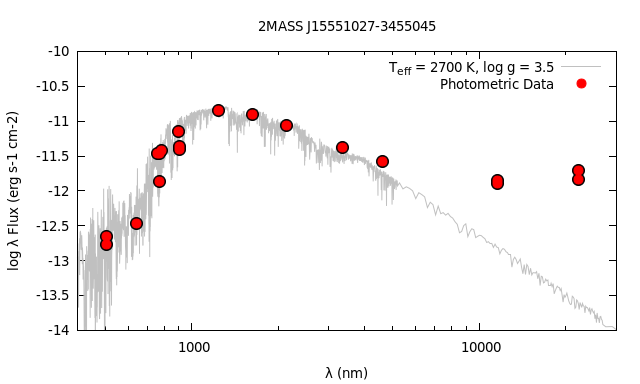}
\includegraphics[width=9cm]{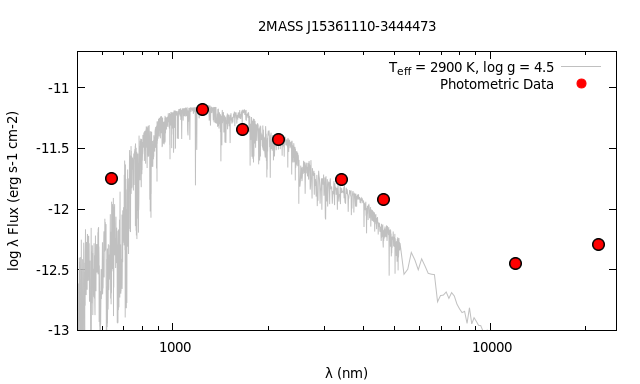}
\includegraphics[width=9cm]{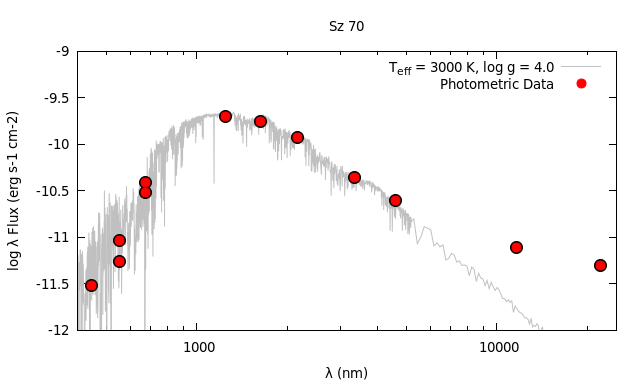}
\includegraphics[width=9cm]{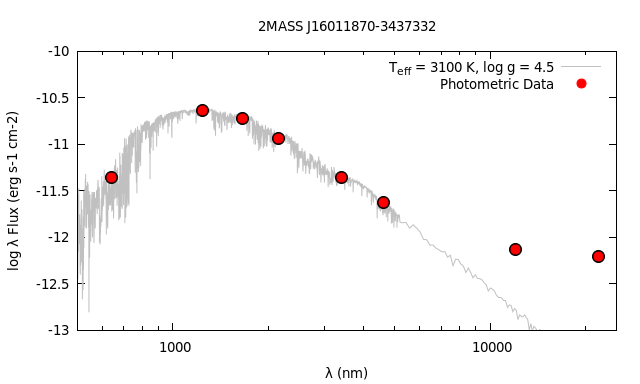}
    \caption{BT-Settl models (in grey) with the photometric data (red dots) for our accretors.}
    \label{fig:seds-accretors}
\end{figure*}

\begin{figure*}[!ht]
\includegraphics[width=9cm]{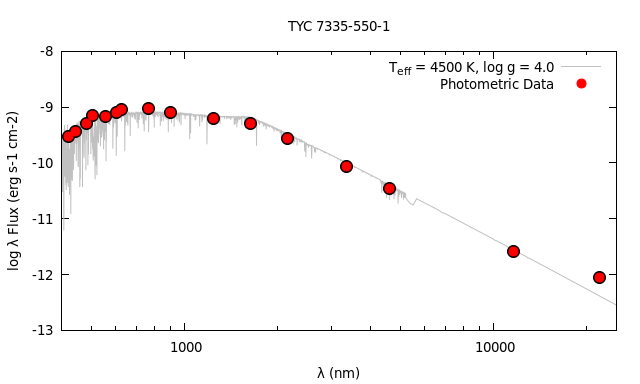}
\includegraphics[width=9cm]{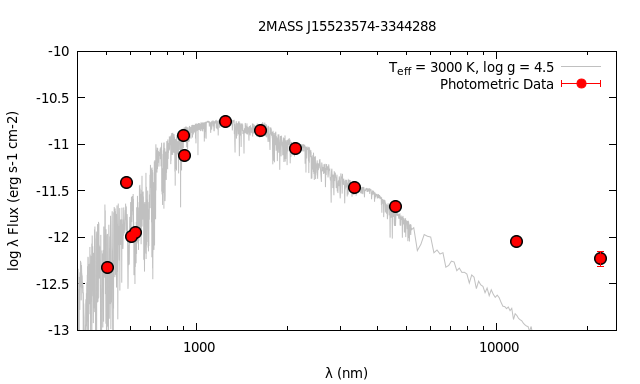}
\includegraphics[width=9cm]{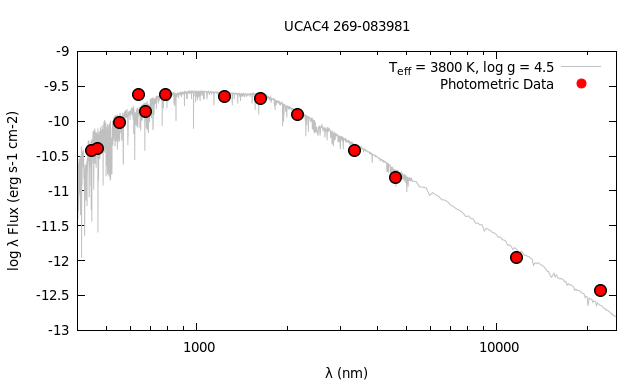}
\includegraphics[width=9cm]{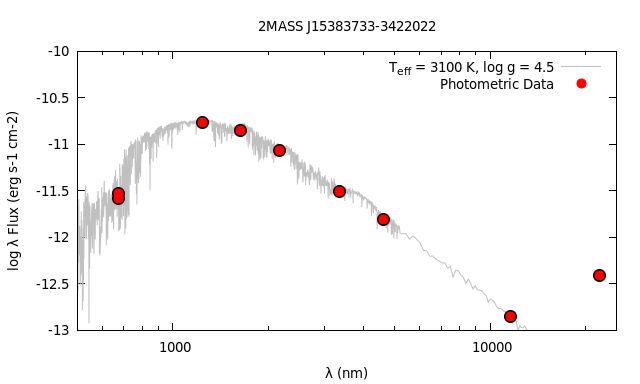}
\includegraphics[width=9cm]{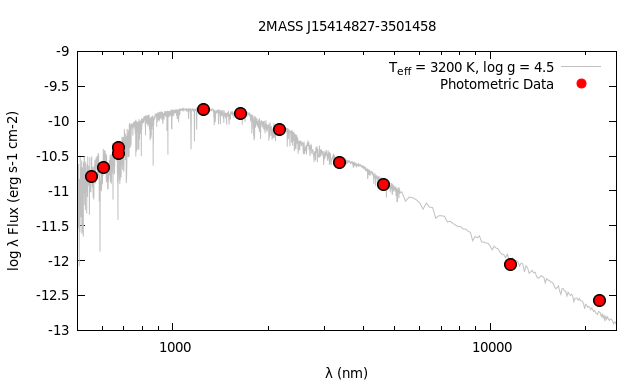}
\includegraphics[width=9cm]{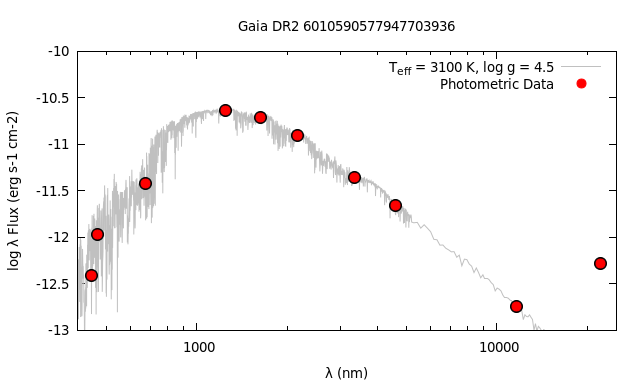}
\includegraphics[width=9cm]{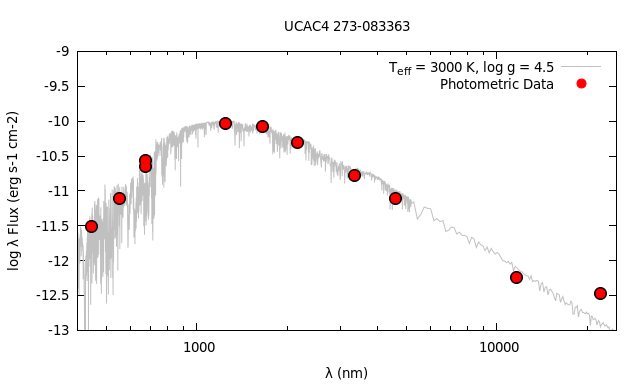}
\includegraphics[width=9cm]{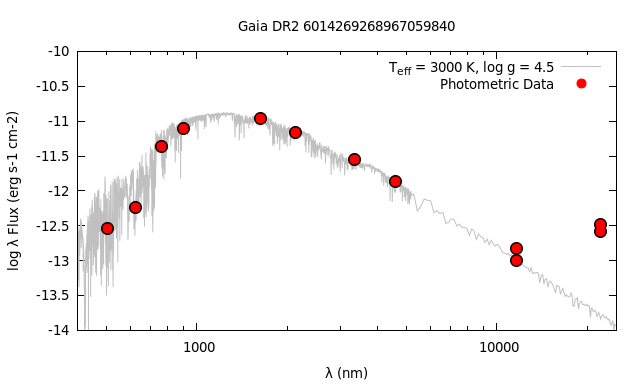}
    \caption{BT-Settl models (in grey) with the photometric data (red dots) for our chromospherically-dominant targets.}
    \label{fig:seds-non-accretors}
\end{figure*}

In the following, we will discuss further qualities of our targets, which are mainly based on the data available in the literature in connection with the targets analyzed in this work.

\subsection{Spectral energy distributions / Circumstellar disks}

\label{disk_seds}

For all our objects, we also investigated whether there are hints of continuum flux excess suggestive of circumstellar disks. To this aim, we extracted their SEDs from literature which are collectively exhibited in Figs.~\ref{fig:seds-accretors} and \ref{fig:seds-non-accretors}. For this work, we only concentrate on the morphology and trends of the SEDs of our targets, as well as their near- to mid-infrared photometric data (published by 2MASS and WISE surveys). For generating the SEDs, we have used the following WISE filters: $W1$ (3.4 microns), $W2$ (4.6 microns), $W3$ (12 microns), $W4$ (22 microns). In a parallel paper (Majidi et al. in prep), we will study the variability of these stars and model their disks.   

The photometric data for all four accretors significantly deviate from their BT-Settl spectral model (based on their $T_{\rm eff}$, log g, and zero metallicity) in $W3$ and $W4$ filters (with the average flux errors of 5e-17 $W.m^{-2}$ and 1.7e-16 $W.m^{-2}$ respectively). This trend can be observed for our less massive, stronger accretors \masszff \hspace{0.02cm} and \massfst \hspace{0.02cm} in all four WISE filters ($W1$, $W2$, $W3$, and $W4$). 
According to \citet{s-a14}, the morphology of the SEDs of all our four accretors in addition to \masstee\ is compatible with objects surrounded by full disks. This is further confirmed by the disk categorization of \citet{bredall20} based on $K_s - W3$ and $K_s - W4$ magnitudes for Lupus dippers, Lupus YSOs, Upper Scorpius and Taurus members. Hence, also according to \citet{bredall20}, all our four accretors in addition to \masstee\ are surrounded by a full disk.  Note, however, that the ``valley'' around $W3$ in the SED of \massfst\ is typical of those seen in transitional disks.

For the rest of our targets, we have two categories of circumstellar disks based on the morphology of their SEDs further approved by their $K_s - W3$ and $K_s - W4$ magnitudes: i) Evolved disks, which are characterized by only $W4$ excess with respect to the theoretical BT-Settl model, and are evident in the SEDs of \massztt, \gaiants, and \gaiaefz\ (Fig. \ref{fig:seds-non-accretors}), ii) Debris disks, which are characterized by little to no mid-infrared excess, and is evident in the SEDs of TYC 7335-550-1, UCAC4 269-083981, \massffe, and UCAC4 273-083363 (Fig. \ref{fig:seds-non-accretors}).

\begin{table*}[!ht]
        \centering
        \caption{Disk categorization of all our targets, in addition to their reddest colors available in the 2MASS and WISE catalogs.}
        \begin{minipage}{\textwidth}
        \centering
        \begin{tabular}{lcccccccccccr} 
                \midrule
                Name & $K_s - W3$ & $K_s - W4$ & \citet{bredall20} & \citet{s-a14}\\                                 & mag &mag & Disk type & SED/Disk type \\ \midrule
2MASS J15383733-3422022 & 0.75 & 3.93 & Evolved disk & \\
Sz 70 & 2.28 & 3.9 & Full disk & Full disk    \\
TYC 7335-550-1 & 0.20 & 1.14 & Debris disk &	\\      
2MASS J15361110-3444473	& 2.70 & 5.04 & Full disk &  Full disk			\\     
2MASS J15523574-3344288 & 2.69 & 4.31 & Full disk & Full disk  \\
2MASS J15551027-3455045 &3.24 & 5.7 & Full disk & Full disk\\
2MASS J16011870-3437332 & 2.18 & 4.09 & Full disk &  Full disk\\
UCAC4 269-083981 & 0.13 & 1.06 & Debris disk & \\
\textit{Gaia} DR2 6010590577947703936 & 0.61 & 3.79 & Evolved disk & \\
2MASS J15414827-3501458 & 0.39 & 1.16 & Debris disk & \\
UCAC4 273-083363 & 0.4 & 1.86 & Debris disk & \\
\textit{Gaia} DR2 6014269268967059840 & 0.89 & 3.58 & Evolved disk &  \\    \hline
        \end{tabular}

        \raggedright
        \textbf{Notes.} The overall SED of \massfst\ may be affected by a possible unresolved M8-type companion.
        \end{minipage}
        \label{table:age}
\end{table*}

\subsection{High accretion in the low-mass regime}
Deriving \Macc ~for the lowest mass accretors is relevant for the studies of disk evolution. There is growing evidence 
of a change in the slope of the \Mstar--\Macc\  relationship for YSOs with ages of 2-3\,Myr at \Mstar$<$0.2\,\Msun\ (\citealt{manara17b} and \citealt{alcala17}, and see Fig.~\ref{fig:mass-accretion}). Such a break could be related to a faster disk evolution at the low-masses (e.g. \cite{vb2009}). To verify this,  the \Macc--\Mstar\ relationship needs to be sampled at much lower \Mstar\ and \Macc\ 
values than done so far.
 
Our target \masszff\ is one of the lowest mass accretors in Lupus (see Fig.~\ref{fig:hrdiagram}). With \Mstar$=$0.02\,\Msun, 2MASS\,J16085953-3856275 is the accretor with comparable mass reported in the previous Lupus studies \citep{alcala17, alcala19}. Considering the very low mass of this YSO, its accretion rate \Macc$\sim$10$^{-11}$\,\Msun/yr \citep[][]{alcala19} is relatively high.
Yet the \Macc\ value for \masszff\ is about an order of magnitude higher (see Fig.~\ref{fig:mass-accretion}); hence, it is one of strongest accretors in Lupus in the mass range ~0.02--0.03\Msun, i.e. close to the planetary mass regime.  
From modeling of a shock at the surface of a planetary-mass object, \cite{aoyama21} have predicted much higher \Lacc\ values than what the scaling \Lacc--\Ll\ relations for stars would predict. The relationships by these authors would yield an even higher \Macc\ value, 
almost an order of magnitude higher than our estimate.
This object falls above the model prediction by \citet{vb2009},  in contrast with the idea of faster disk evolution at very 
low masses. However, statistics are still rather poor at this mass regime for a firm conclusion. 

Other very low-mass YSOs, companions to T Tauri stars, have been found to exhibit similar, or even higher rates 
of mass accretion \citep[][see Fig.~\ref{fig:mass-accretion}]{betti22, zhou2014}. To explain the very high levels of accretion observed 
in such sub-stellar and planetary-mass companions, \citet{staher15} modeled the accretion onto very low-mass objects that formed by the fragmentation of the disk around the hosting star. During the early evolution the individual disks of sub-stellar companions, including those at the planetary-mass regime, accrete additional material from the gas-rich parent disk, hence, their disks are more massive and their accretion rates are higher than if they were formed in isolation. Therefore, these very low-mass objects have disk masses and accretion rates that are independent of the mass of the central object and are higher than expected from the scaling relation \Macc~$\propto$~$M_{\star}^2$ of more massive YSOs. These models predict that \Macc\ is independent of \Mstar. 

Using \textit{Gaia} DR3, we have investigated whether \masszff\ might be a wide companion of another star, but it is an isolated object. Hence,
the high mass accretion rate cannot be explained in terms of the \citet{staher15} scenario. Due to its intrinsic faintness, \masszff\ would be an interesting target to be followed up by CUBES, which is a next-generation spectrograph suitable for investigating fainter, low-mass accreting YSOs \citep{alcala-cubes}.  

\subsection{Possible wide companions}
\label{wide_companions}

While studying the kinematic properties of the targets, we also noticed that a few of our targets and core members of the Lupus I share similar kinematic properties, and can be considered as wide companion candidates. These wide companion candidates are presented in Table \ref{similar-targets} and Table \ref{similar-core}, divided into two categories of candidates studied in this work and the Lupus I core members. In order to understand whether two objects with similar kinematic properties are gravitationally bound, we calculated their total velocity difference ($
\Delta v$) and compared it with the maximum total velocity difference ($\Delta v_{max}$) as a function of projected separation
between the two binary components, suggested by \citet{andrews17}. If $\Delta v$ exceeds $\Delta v_{max}$, we do not expect the two targets to be gravitationally bound. It should be noted, however, that the theoretical maximum velocity difference modeled by \citet{andrews17} is only for binaries of total mass 10 $M_\odot$ in circular orbits. We summarize our results on identifying wide companions candidates in the Lupus I cloud as follows:

\begin{table*}[!ht]
        \centering
        \caption{Kinematic properties of the Lupus I members from this work (measurement errors are displayed in parenthesis).}
        \resizebox{\textwidth}{!}{\begin{tabular}{lcccccccccccr} 
                \midrule
                Name & $\alpha$ (J2000) & $\delta$ (J2000) & $\varpi$ & $\mu_{\alpha*}$  & $\mu_\delta$ & RV & Age & $\Delta V$ & $\delta_{\Delta V}$ & S \\
                & (h:m:s) & (d:m:s) & (mas) & (mas/yr) & (mas/yr) & (km/s) & (Myr) & (km/s) & (km/s) & (\arcsec)\\
                \midrule
Sz 71/GW LUP$^{*}$ &	                                         											      15 46 44.73 &	--34 30 35.68 &	6.41(0.06) &	--14.03(0.10)	& --23.36(0.07) &  --3.30(1.90) & 2.0 &  6.07 & 3.24 & 32.32\\    
Sz 70 & 15 46 42.99 & --34 30 11.55 & 6.09(0.21) & --12.58(0.39) & --22.16(0.25) & 1.1(2.6) & 0.5 
\\
                      
                \midrule
2MASS J15361110-3444473 & 15 36 11.09 & --34 44 47.82 & 5.83(0.29) & --13.56(0.29) & --20.21(0.23) & 6.9(2.6) & 9.77 & 4.72 & 3.47 & 16.28 \\

TYC 7335-550-1 &  15 36 11.55 & --34 45 20.54 & 6.26(0.07) & --13.93(2.43) & --19.51(1.01) & 2.6(2.0) & 3.55 \\

\midrule

              \end{tabular}}
\label{similar-targets}
\raggedright

 $^{*}$ RV obtained by \citet{frasca17}. \\  

\end{table*}

\begin{table*}[!htb]
        \centering
        \caption{Core members of Lupus I sharing similar kinematic properties (measurement errors are displayed in parenthesis).}
        \resizebox{\textwidth}{!}{\begin{tabular}{lcccccccccccr} 
                \midrule
                Name & $\alpha$ (J2000) & $\delta$ (J2000) & $\varpi$ & $\mu_{\alpha*}$  & $\mu_\delta$ & RV & Age &$\Delta V$ & $\delta_{\Delta V}$ & S \\
                & (h:m:s) & (d:m:s) & (mas) & (mas/yr) & (mas/yr) & (km/s) & (Myr)& (km/s) & (km/s) & (\arcsec)\\
                \midrule
Sz 65/V$^{*}$ IK Lup$^{*}$	&                                          												   15 39 27.77 &	--34 46 17.21 &	6.44(0.05) &	--13.27(0.12) &	--22.24(0.07) & --2.70(2.00) & 1.9 & 5.26 &2.69 &6.41 \\
Sz 66$^{*}$ &	                                           												   15 39 28.28 &	--34 46 18.09 &	6.36(0.09) & --13.60(0.19) &	--21.56(0.12) & 2.40(1.80) & 3.9\\
                
\midrule                

Sz 68/HT LUP A-B$^{*}$	& 																					   15 45 12.87 &	--34 17 30.65 &   6.49(0.06) & 	--13.63(0.13) & 	--21.60(0.08) & --4.30(1.80) & 0.5 & 6.30 & 4.30 & 2.82 \\

CD-33 10685C/HT Lup C$^{**}$ &            																   15 45 12.67&	--34 17 29.37 &   6.55(0.19)& 	--15.43(0.22) & 	--20.27(0.15)& 1.2(3.9) & --\\

\midrule
              \end{tabular}}
\label{similar-core}
\raggedright
$^{*}$ RV and age obtained by \citet{frasca17}.
\\ $^{**}$ RV for this target is adopted from the optimal RV calculated by BANYAN $\Sigma$, considering HT Lup C is a member of UCL. 
\end{table*}

\begin{figure}[!htb]
\centering
\includegraphics[width=9cm]{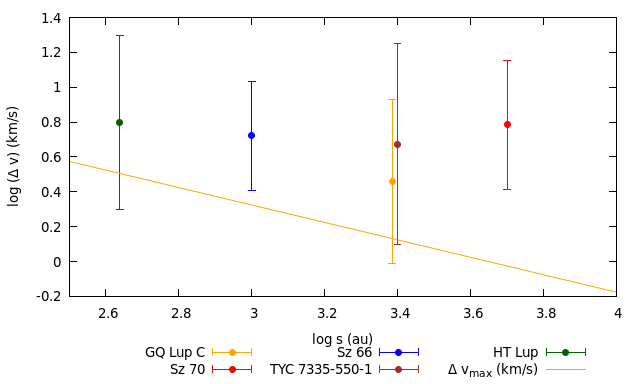}
\caption{Log-log plot of total velocity difference $\Delta v$ (km/s) versus projected separation s (au) for the wide companion candidates analyzed in this work, in addition to the genuine wide companions GQ Lup and GQ Lup C. $\Delta v_{max}$ (km/s) (orange line) indicates the maximum total velocity difference that bound binaries with a total mass equal to 10 $M_{\odot}$ in circular orbits can possess \citep{andrews17}. Each point is marked as one of the wide companion candidates involved. For the detailed information, see Tables \ref{similar-targets} and \ref{similar-core}.}
\label{fig:companions}
\end{figure}

\textbf{Sz 70 and Sz 71 --} Same as the GQ Lup triple system \citep{alcala2020}, Sz 70 and Sz 71 (GW Lup) are located on the main filament of Lupus I. Sz 70 lies at a separation of 32.32 arcseconds from GW Lup, and in between these objects lies the X-ray source [KWS97] Lupus I 37 \citep[][]{krautter1997} at a separation of 24.23 arcseconds from Sz 70. We conducted a chance projection study in \citet[][Appendix E]{alcala2020}, which was focused on understanding how probable it is to find a field object around a genuine member of Lupus I, lying on the same filament where GQ Lup stellar system and Sz 70/Sz 71 are located. The linear density of this filament is 0.0024 objects/arcsec, or an average object separation of 418 arcsec, which is 13 times the observed separation between Sz 70 and Sz 71. As exhibited in Fig. \ref{fig:companions}, Sz 70 and Sz 71 do not qualify as gravitationally bound stars, but we would like to emphasize that the test proposed by \citet{andrews17} is only valid for gravitationally bound binaries, and not systems of higher multiplicities (if this is the case for this stellar system). Hence, we would consider this case as a wide companion candidate that cannot be confirmed or ruled out according to the available information.

\textbf{TYC 7335-550-1 and 2MASS J15361110-3444473 --} As discussed in Sect. \ref{results}, 2MASS J15361110-3444473 might be an unresolved binary, composed of an M6 (VIS spectrum) and an M8 (NIR spectrum) star. The RV calculated for this target based on the ROTFIT code is obtained by cross-correlations conducted on the VIS spectrum of this target, which is also used for calculating the maximum velocity difference between TYC 7335-550-1 and 2MASS J15361110-3444473. As exhibited in Fig. \ref{fig:companions}, the two objects can be gravitationally bound. However, TYC 7335-550-1 has an age of $\sim$ 4 Myr and 2MASS J15361110-3444473 an age of $\sim$ 9 Myr, which states the two stellar systems are probably not coeval. Also, unlike TYC 7335-550-1, we could not determine whether 2MASS J15361110-3444473 is a member of Lupus I due to many uncertainties explained earlier. Hence, any further comments on its physical association with TYC 7335-550-1 would be misleading and inconclusive.

\textbf{Sz 65 and Sz 66 --} At a separation of 6.45 arcseconds, with $\Delta V$ = 5.26$\pm$2.69 km/s, Sz 65 and Sz 66 (although coeval) according to the test suggested by \citet{andrews17} are not gravitationally bound. There are no other objects located in a close separation with respect to either Sz 65 or Sz 66. Hence, we rule out the possibility of Sz 65 and Sz 66 as wide companion candidates. 

\textbf{HT Lup A-B-C --} This stellar system is located in an over-crowded region on the same filament of Lupus I as GQ Lup stellar system. In \textit{Gaia} DR2 catalog, HT Lup A and B are not resolved separately, hence we assume the central star to be Sz 68 (or HT Lup A), composed of two unresolved stars, and adopt its stellar characteristics from \citet{frasca17}. As genuine members of Lupus I, we assume all the components of this triple system to have an age consistent with the other bona fide members of Lupus I ($\leq$ 2 Myr), and hence, to be coeval. However, the RVs used here should be taken with caution, both because HT Lup A-B are not resolved, and also because we have adopted the optimal RV calculated by BANYAN $\sigma$ for HT Lup C considered as a member of UCL. With a separation of 2.82 arc seconds, we have shown in Fig. \ref{fig:companions} that as expected, this triple system is possibly gravitationally bound.    

We thus conclude that the possibility of Sz 70 \& Sz 71 being wide companions is rather low and for TYC 7335-550-1 \& 2MASS J15361110-344447, follow-up studies on 2MASS J15361110-344447 are required. As for the previously identified members of Lupus I, we understood that Sz 65 and Sz 66 are not gravitationally bound, and HT Lup A-B-C are the components of a triple system.

\section{Conclusion}

The main conclusions of this paper can be summarized as follows:

\begin{itemize}
    \item Out of the 12 objects fully characterized in this work, ten are recognized as genuine members of Lupus~I, and two remain ambiguous in terms of stellar properties. 
    \item Out of the ten members of Lupus I analyzed in this work, three were recognized to be accretors (Sz 70, \masszff, and \massttt), and Sz 70 and \masszff\ are likely to be surrounded by full disks. \masszff\ is among the least massive accretors discovered so far in the Lupus complex, formed in full isolation and is an off-cloud member of Lupus I.   
    \item All of the three off-cloud targets included in our program turned out to be genuine members of Lupus~I. These targets are \masstee, \masszff, and \massttt, with \masszff\ and \massttt\ actively accreting matter, and \masstee\ mildly accreting matter. Further investigation in this area may 
    reveal a diffused population of M dwarfs close to the main filament of Lupus~I. We thus would like to acknowledge that this work also contributes to revealing the diffused populations of M-dwarfs around the Lupus cloud by \citet{comeron2008}.  
    \item Although the sample studied in this work is small, we proved that many interesting targets in young star forming regions can escape \Ha\ surveys due to various reasons. Hence, using the kinematic properties of candidate YSOs can play a key role in identifying the genuine members of the young stellar associations. This is specifically true for genuine members such as TYC~7335-550-1 that have \Ha\ in absorption, and hence would not appear in \Ha\ surveys. 
    
    \item We have identified a plausible binary system among the targets analyzed in this work, namely, TYC 7335-550-1 and 2MASS J15361110-3444473. It is noteworthy, however, that 2MASS J15361110-3444473 might be an unresolved binary, and its kinematic properties (especially RV) should be revised with next-generation spectrographs (due to its intrinsic faintness).

    \item All the above points considered, we conclude that characterizing only a small portion of our sample has proved to have a high success rate for discovering the new members of Lupus~I. This shows that the spectroscopy of our entire sample of 43 objects could have resulted in a far more solid investigation of the region in terms of determining the disk fraction, stellar properties, and the number of new members of Lupus~I.
\end{itemize}

\begin{acknowledgements}

FZM is grateful to Eugene Vasiliev for fruitful discussions on how to use \textit{Gaia} catalogs. AFR is grateful to Giovanni Catanzaro for helping us with the analysis of  TYC 7335-550-1. FZM is funded by "Bando per il Finanziamento di Assegni di Ricerca Progetto Dipartimenti di Eccellenza Anno 2020" and is co-funded in agreement with ASI-INAF n.2019-29-HH.0 from 26 Nov/2019 for "Italian participation in the operative phase of CHEOPS mission" (DOR - Prof. Piotto). A.B. acknowledges partial funding by the Deutsche Forschungsgemeinschaft Excellence Strategy - EXC 2094 - 390783311 and the ANID BASAL project FB210003. JMA, AFR, CFM, KBI and ECO acknowledge financial support from the project PRIN-INAF 2019 ``Spectroscopically Tracing the Disk Dispersal Evolution'' (STRADE).
CFM is funded by the European Union under the European Union’s Horizon Europe Research \& Innovation Programme 101039452 (WANDA). This work has also been supported by the PRIN-INAF 2019 "Planetary systems at young ages (PLATEA)" and ASI-INAF agreement n.2018-16-HH.0. Views and opinions expressed are however those of the author(s) only and do not necessarily reflect those of the European Union or the European Research Council. Neither the European Union nor the granting authority can be held responsible for them.

This work has made use of data from the European Space Agency (ESA) mission
\textit{Gaia} (https://www.cosmos.esa.int/gaia), processed by the \textit{Gaia}
Data Processing and Analysis Consortium (DPAC,
https://www.cosmos.esa.int/web/gaia/dpac/consortium). Funding for the DPAC
has been provided by national institutions, in particular, the institutions
participating in the \textit{Gaia} Multilateral Agreement.

This research has made use of the SIMBAD database and Vizier services, operated at CDS, Strasbourg, France.
This research has made use of the services of the ESO Science Archive Facility.

Finally, we would like to thank the anonymous referee who also contributed to this paper with his/her valuable comments.
\end{acknowledgements}

\newpage
\begin{appendix}
\normalsize

\section{Candidate members of Lupus I}
\label{candidate-list}

As we explained in Sect. \ref{sec:obs}, we proposed 43 objects to be observed with X-Shooter. Twelve out of these 43 objects were observed during a filler program, and in this work we fully characterized them. The rest of our targets in this sample that were not observed are listed in Table \ref{table:extra-obj}. Among these targets, only 2MASS J15464664-3210006 \citep{eisner2007} is partly characterized, and 20 objects are identified as candidate YSOs using \textit{Gaia} DR2 \citep{zari18}.

\begin{table*}[!h]
        \centering
        \caption{Astrometric properties of the candidate Lupus I members that were not observed by X-Shooter, with their errors in parentheses.}
        \begin{minipage}{\textwidth}
        \begin{tabular}{lccccccccr} 
                \midrule
                Name & $\alpha$ (J2000) & $\delta$ (J2000) & $\varpi$ & $\mu_{\alpha*}$  & $\mu_\delta$ & \textit{J}  \\
                & (h:m:s) & (d:m:s) & (mas) & (mas/yr) & (mas/yr) & (mag)  \\
                \midrule
2MASS J15464664-3210006$^a$ & 15 46 46.64 & --32 10 00.62 & 7.05(0.021) & --19.47(0.023) & --23.76(0.014) & 11.22 \\
Gaia DR2 6013000844869745664 & 15 39 24.47 & --35 58 50.88 & 6.62(0.039) & --18.00(0.081) & 	--22.23(0.057) & 10.11 \\
Gaia DR2 6013065853493820416$^b$ & 15 43 15.62 & --35 39 38.18 & 6.88(0.015) & --17.68(0.018) & --24.51(0.012) &  10.20   \\
 Gaia DR2 6011737574730221568$^c$ & 15 50 46.50& --34 22 38.49 & 6.69(0.019) & --16.20(0.020) & --22.52(0.015) &  10.74\\
 Gaia DR2 6012258330925877632$^d$ & 15 53 36.13& --33 31 02.60 & 6.92(0.016) & --16.97(0.018) & --24.57(0.016) & 10.75\\
Gaia DR2 6039383622075982848$^e$ & 15 57 09.76& --32 04 33.91 & 6.72(0.02) & --14.24(0.023) & --23.58(0.015) & 10.56\\
 Gaia DR2 6011518462675791872$^f$ & 15 48 13.16& --35 43 31.08 & 6.62(0.023) & --16.65(0.028) &  --24.31(0.023) & 11.48\\
Gaia DR2 6011797738632729216$^g$ & 15 57 20.96& --35 00 01.21 & 6.71(0.027) & --16.29(0.033) & --24.21(0.024) & 11.65\\
Gaia DR2 6014049985115937408 & 15 34 59.21&	--34 58 16.16 & 6.83(0.097) & 	--17.76(0.16) & --24.03(0.11) & 12.16\\
  Gaia DR2 6014830844535625344$^h$ & 15 47 58.08& --33 46 59.53 & 6.84(0.027) & --17.73(0.031) & --24.48(0.025) & 11.31\\
  Gaia DR2 6014224051546189568 & 15 34 42.05&	--34 17 48.09 & 6.66(0.098)& 	--17.36(0.134) 	&--23.67(0.094) & 11.94 \\
   Gaia DR2 6009936093645659136 & 15 43 49.43&	--36 48 38.64 & 6.94(0.13) & 	--20.45(0.28) & 	--22.89(0.19) & 10.92\\
    Gaia DR2 6039633559115225344$^i$ & 15 52 59.02& --31 38 33.57& 6.59(0.03) & --18.34(0.036)& --22.89(0.029) & 11.93 \\
  Gaia DR2 6013187040287810944$^j$ & 15 37 53.31& --35 55 12.42 & 6.74(0.027) & --17.9(0.03) &--24.08(0.024) & 11.95\\
   Gaia DR2 6016139332082870272 & 15 39 25.88&	--32 10 04.68& 6.42(0.40)& 	--20.32(0.54)& 	--23.65(0.37) & 10.81\\
    Gaia DR2 6013126738951338624$^k$ & 15 43 28.48& --35 17 27.40 & 6.77(0.032) & --17.67(0.035) & --24.48(0.022) & 11.91\\
     Gaia DR2 6013190201383772288 & 15 37 53.00&	--35 52 28.70 & 6.75(0.055) & 	--19.08(0.13) & --22.62(0.087) & 12.22\\
        Gaia DR2 6013077192207599232$^m$ & 15 43 11.42&  --35 26 34.43& 6.78(0.032) & --17.32(0.034)& --24.29(0.025) & 11.82\\
         Gaia DR2 6015181897983193728$^m$ & 15 51 57.84& --33 29 33.17 & 6.74(0.032) &--16.22(0.039) & --22.37(0.026) &  12.03\\
  Gaia DR2 6014590429442468096$^m$ & 15 45 06.91& --35 06 21.73 &6.99(0.036)& --16.97(0.042) & --23.09(0.029) & 11.82\\
    Gaia DR2 6009995742152335232$^m$ & 15 44 26.97& --36 25 42.75& 6.52(0.034) & --18.30(0.043)& --23.21(0.031) & 11.82\\
     Gaia DR2 6011607694917034112$^m$ & 15 50 00.76& --35 29 19.71 & 7.23(0.044) & --20.18(0.052) & --25.32(0.034) & 12.37\\
      Gaia DR2 6011695690208264320$^m$ & 15 47 59.03& --34 56 38.36& 6.99(0.06) & --17.93(0.069)&  --25.07(0.045) & 12.69\\
       Gaia DR2 6011261726715424128 &15 50 29.19&	--36 25 11.80& 6.92(0.11)& 	--17.08(0.23) &--23.52(0.16) & 13.32\\
        Gaia DR2 6015222957871475584 & 15 48 46.12&	--33 18 35.48& 6.69(0.13)& 	--19.21(0.26) & 	--23.77(0.17) & 13.77\\
        Gaia DR2 6013030875279571328 &15 41 55.22&	--35 59 35.36 & 6.97(0.12)& 	--17.12(0.24)&--25.52(0.14) & 13.17\\
          Gaia DR2 6014112107523072640$^m$ &15 34 35.79& --34 36 01.54 & 6.88(0.084) & --16.89(0.087) & --24.841(0.066) & 13.14\\
           Gaia DR2 6012977136650130560$^m$ & 15 39 48.47& --36 13 48.07 & 6.94(0.10) & --20.069(0.11) & --23.61(0.069) & 12.81\\
            Gaia DR2 6015141830223216640 & 15 50 19.17&	--33 50 07.12& 6.84(0.15) & --17.29(0.29) & --26.46(0.19) & 13.92\\ 
            Gaia DR2 6011581856393988352$^n$ & 15 48 06.26 & --35 15 48.15 & 6.05(0.07) & --12.22(0.084)& --21.04(0.057) & 10.56\\
             Gaia DR2 6016191485871670400 & 15 38 35.63& --32 02 37.66& 6.53(0.26) & --18.90(0.39)& 	--23.38(0.28) & 14.35\\
\midrule
        \end{tabular}
\\    $^a$ 2MASS J15464664-3210006 is an M2, T Tauri star \citep{eisner2007}.\\
    $^b$ aka UCAC4 272-080482, this target is a YSO candidate \citep{zari18}.\\
    $^c$ aka UCAC4 279-083370, this target is a YSO candidate \citep{zari18}.\\
    $^d$ aka UCAC4 283-086052, this target is a YSO candidate \citep{zari18}.\\
    $^e$ aka RX J1557.1-3204A, this target is a YSO candidate
    \citep{zari18}.\\
    $^f$ aka UCAC4 272-081081, this target is a YSO candidate \citep{zari18}.\\
    $^g$ aka UCAC4 275-083957, this target is a YSO candidate \citep{zari18}.\\
    $^h$ aka UCAC4 282-082547, this target is a YSO candidate \citep{zari18}.\\
    $i$ aka UCAC4 292-084899, this target is a YSO candidate \citep{zari18}.\\
    $^j$ aka UCAC4 271-080669, this target is a YSO candidate \citep{zari18}.\\
    $^k$ aka UCAC4 274-080590, this target is a YSO candidate \citep{zari18}.\\
    $^l$ aka UCAC4 274-080590, this target is a YSO candidate \citep{zari18}.\\
    $^m$ This target is a YSO candidate \citep{zari18}.\\
    $^n$ aka UCAC4 274-081081, this target is a YSO candidate \citep{zari18}.\\    \label{table:extra-obj}
\end{minipage}
\end{table*}

\section{Age estimation and isochrones}
\label{appen:age}

For estimating the age of our targets we used multiple isochrones for the reasons explained in Sect. \ref{sec:membership-criteria}. In this Appendix, we present the ages of our targets using various isochrones. We repeat that the ages estimated for all our targets were overestimated by PARSEC models in comparison with all the other models with a considerable gap. We thus decided to remove the results achieved by the PARSEC models to avoid confusion. This is, however, a well-known problem of PARSEC isochrones that they overestimate the age of cool stars, and all our targets fall in this category. 

\begin{table*}[!h]
        \centering
        \caption{Ages of our targets estimated using various isochrones. The ages are all in Myr.}
        \begin{minipage}{\textwidth}
        \centering
        \begin{tabular}{lcccccccccccr} 
                \midrule
                Name & Dartmouth  & Dartmouth & MIST & Baraffe \\                                 & std & mag & & models \\ \midrule
2MASS J15383733-3422022 & 11 & 20 & 12.6 & 10.7\\
Sz 70 & $<$1 & 1 & $<$0.25 & 0.5   \\
TYC 7335-550-1 & 3 & 5 & 3.5 & 3.55	\\      
2MASS J15361110-3444473	& 9 & 20 & 9& 9.77 			\\     
2MASS J15523574-3344288 & 8 & 13 & 8 & 6.3  \\
2MASS J15551027-3455045 & - & - & -\footnote{None of the three isochrones used here were able to reproduce the stellar parameters of this target due to its dimness.} & 1.7 \\
2MASS J16011870-3437332 & 9.5 & 14 & 9.5 & 9.55 \\
UCAC4 269-083981 & 4.5 & 8 & 3.5 &4.2 \\
\textit{Gaia} DR2 6010590577947703936 & 8 & 14 & 8 &8.8 \\
2MASS J15414827-3501458 & 2.5 & 3 & 1.78 &1.82 \\
UCAC4 273-083363 & 4.5 & 8 & 3.5 &3.63 \\
\textit{Gaia} DR2 6014269268967059840 & 8 & 13 & 8 & 6.46 \\    \hline
        \end{tabular}
        \end{minipage}
        \label{table:age}
\end{table*}

\section{2MASS J15361110-3444473}
\label{app:notes-targets}

\begin{figure}[H]
\centering
\includegraphics[width=9cm]{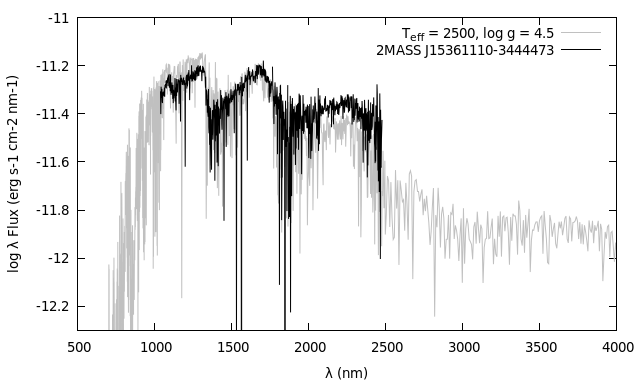}
\caption{Flux-calibrated, extinction-corrected NIR spectrum of \massfst\ (in black) with its BT-Settl model ($T_{eff}$ = 2500 K and log $g$ = 4.5, in grey).}
\label{fig:sed-nir-2mass473}
\end{figure}

2MASS J15361110-3444473 is an M5.5 star according to its VIS spectrum (as we quantitatively indicated) and an M8 star based on its NIR spectrum (based on the fitting done with the BT-Settl model $T_{\rm eff}$ = 2500 K and log $g$ = 4.5, as exhibited in Fig. \ref{fig:sed-nir-2mass473}), with a total extinction of $A_V$ = 1.75 mag. All the spectral typing and analysis that we have performed in this paper are based on the VIS spectrum of this target, especially the ROTFIT results are all based on the VIS spectrum. Hence, although we keep our analysis limited to the spectroscopy conducted on the VIS spectrum, we would like to emphasize that the possibility of this target being an unresolved binary (composed of two M dwarfs) with SpTs of M5.5 and M8 is viable. Considering the available data, we also cannot rule out the possibility that the star is heavily spotted instead of being a binary.

\section{Updates with \textit{Gaia} DR3}

As stated in Sect. \ref{sec:obs}, we used the \textit{Gaia} DR2 catalog to select our targets. Very recently, \textit{Gaia} DR3 \citep{gaiaedr3} became public and gave us the opportunity to check the catalog for any possible changes or updates on the kinematic or stellar properties of our objects analyzed in this work. We did not find any considerable difference between the kinematic properties reported in both catalogs. However, we report the highlights of our search using these two catalogs in the following:

\textbf{TYC 7335-550-1 --} as obtained in this work, for TYC 7335-550-1 we obtained $T_{\rm eff}$ = 4488 K, while in both \textit{Gaia} DR2 and \textit{Gaia} DR3 its reported temperature is 5000 K. The reported RV for TYC 7335-550-1 in \textit{Gaia} DR2 is 1.20$\pm$1.65 km/s, which is better constrained than the RV we report here (2.6$\pm$2.0 km/s). As the wide companion candidate of \massfst, we recalculated their $\Delta v$ using the \textit{Gaia} DR3 kinematic properties of TYC 7335-550-1, and it resulted in $\Delta v$ = 5.34$\pm$3.30 (km/s) which is consistent with the previous $\Delta v$ = 4.72$\pm$3.47 (km/s). For both of these calculations, we use the RVs calculated by ROTFIT. 

\textbf{Sz 70 --} has a high RUWE in both catalogs (4.86), but we saw no signs of binarity in the spectrum of Sz 70. Using the kinematic properties of Sz 70 reported in \textit{Gaia} DR3 and those of Sz 71 (which is also updated in \textit{Gaia} DR3), we recalculated their maximum velocity difference, and it resulted in $\Delta v$ = 8.36$\pm$3.24 (km/s), which is consistent with the $\Delta v$ = 6.07$\pm$3.24 (km/s) calculated based on \textit{Gaia} DR2.

\textbf{2MASS J15414827-3501458 --} has a high RUWE (4.198) in both \textit{Gaia} DR2 and \textit{Gaia} DR3 catalogs, but we detected no signs of binarity in the spectrum of the object.

We report that the kinematic properties of all our targets (parallax and proper motions) are consistent within 3$\sigma$ in the two catalogs. Also according to \citet{manara2022}, we do not expect the stellar physical parameters of our core sample to be changed with the astrometry reported in \textit{Gaia} DR3.

\end{appendix}


\begin{thebibliography}{}

\bibitem[\protect\citeauthoryear{Alcal\'a et al.}{2014}]{alcala14}
Alcal\'a, J. M., Natta, A., Manara, C., et al. 2014, A\&A, 561, A2

\bibitem[\protect\citeauthoryear{Alcal\'a et al.}{2017}]{alcala17}
Alcal\'a, J. M., Manara, C., Natta, A., et al. 2017, A\&A, 600, 20

\bibitem[\protect\citeauthoryear{Alcal\'a et al.}{2019}]{alcala19}
Alcal\'a, J. M., Manara, C., France, K., et al. 2019, A\&A, 629, A108

\bibitem[\protect\citeauthoryear{Alcal\'a et al.}{2020}]{alcala2020}
Alcal\'a, J. M., Majidi, F. Z., Desidera, S., et al. 2020, A\&A, 635, L1 

\bibitem[\protect\citeauthoryear{Alcal\'a et al.}{2022}]{alcala-cubes}
Alcal\'a, J. M., Cupani, G., Evans, C., et al.
2022, Exp Astron, in press as part of the Special Issue

\bibitem[\protect\citeauthoryear{Andrews et al.}{2017}]{andrews17}
Andrews, J. J., Chanam\'e, J., Agueros, M. A., et al. 2017, MNRAS, 472, 675

\bibitem[\protect\citeauthoryear{Asensio-Torres et al.}{2019}]{asensio19}
Asensio-Torres, R., Currie, T., Janson, M., et al. 2019, A\&A, 622, A42 

\bibitem[\protect\citeauthoryear{Aoyama et al.}{2021}]{aoyama21}
Aoyama, Y., Marleau, G.-D., Ikoma, M., Mordasini, Ch. 2021, \apj, 917, 30


\bibitem[\protect\citeauthoryear{Baraffe et al.}{2015}]{baraffe15}
Baraffe, I., Homeier, D., Allard, F., \& Chabrier, G. 2015, A\&A, 577,42

\bibitem[\protect\citeauthoryear{Baratella et al.}{2020}]{baratella20b}
Baratella, M., D’Orazi, V., Carraro, G., et al. 2020, A\&A, 634, A34

\bibitem[\protect\citeauthoryear{Beccari et al.}{2018}]{beccari18}
Beccari, G., Petr-Gotzens, M., Boffin, et al. 2018, The Messenger, 173, 17–21 

\bibitem[\protect\citeauthoryear{Benedettini et al.}{2018}]{benedettini18}
Benedettini, M., Pezzuto, S., Schisano, E. et al. 2018, A\&A, 619, 52

\bibitem[\protect\citeauthoryear{Betti et al.}{2022}]{betti22}
Betti, S. K., Follette, K. B., Ward-Duong, K., et al. 2022, ApJL, 935, L18

\bibitem[\protect\citeauthoryear{Biazzo et al.}{2017}]{biazzo17}
Biazzo, K., Frasca, A., Alcal\'a, J. M., et al. 2017, A\&A, 605, A66

\bibitem[\protect\citeauthoryear{Bildsten et al.}{1997}]{bildstenetal1997} Bildsten, L., Brown, E. F., Matzner, C. D., Ushomirsky, G. 1997, \apj, 482, 442

\bibitem[\protect\citeauthoryear{Binks et al.}{2022}]{binks22}
Binks, A. S., Jeffries, R. D., Sacco, G. G., et al. 2022, MNRAS, 513, 5727

\bibitem[\protect\citeauthoryear{Bouvier et al.}{2016}]{bouvier16}
Bouvier, J., Lanzafame, A. C., Venuti, L., et al. 2016, A\&A, 590, A78

\bibitem[\protect\citeauthoryear{Bredall et al.}{2020}]{bredall20}
Bredall, J. W., Shappee, B. J., Gaidos, E., et al. 2020, MNRAS, 496, 3257

\bibitem[\protect\citeauthoryear{Bressan et al.}{2012}]{bressan12}
Bressan, A., Marigo, P., Girardi, L., et al. 2012, MNRAS, 427, 127

\bibitem[\protect\citeauthoryear{Cayrel}{1988}]{cayrel1988}
Cayrel, R., Proceedings of of the Alpbach Summer school, 1988

\bibitem[\protect\citeauthoryear{Choi et al.}{2016}]{choi16}
Choi, J., Dotter, A., Conroy, C., et al. 2016, ApJ, 823, 102


\bibitem[\protect\citeauthoryear{Clough et al.}{2005}]{clough2005}
Clough, S., Shephard, M., Mlawer, E., et al. 2005, JQSRT, 91, 233

\bibitem[\protect\citeauthoryear{Comer\'on}{2008}]{comeron2008}
Comer\'on, F. 2008, Handbook of Star Forming Regions, Volume II, 5, 295

\bibitem[\protect\citeauthoryear{Comer\'on et al.}{2009}]{comeron2009}
Comer\'on, F., Spezzi, L., \& L\'opez Mart\'i, B. 2009, A\&A, 500, 1045

\bibitem[\protect\citeauthoryear{Comer\'on et al.}{2013}]{comeron2013}
Comer\'on, F., Spezzi, L., L\'opez Mart\'i, B., \& Mer\'in, B., 2013, A\&A, 554, A86

\bibitem[\protect\citeauthoryear{Constantino et al.}{2021}]{constantino21}
Constantino, T., Baraffe, I., Goffrey, T., et al. 2021, A\&A, 654, A146 

\bibitem[\protect\citeauthoryear{Dotter}{2016}]{dotter16}
Dotter, A. 2016, ApJS, 222, 8

\bibitem[\protect\citeauthoryear{Dzib et al.}{2018}]{dzib18}
Dzib, S. A., Loinard, L., Ortiz-Le\'on, G. N., et al. 2018, ApJ, 867, 151

\bibitem[\protect\citeauthoryear{Eisner et al.}{2007}]{eisner2007}
Eisner, J. A., Hillenbrand, L. A., White, R. J., et al. 2007, ApJ, 669, 1072


\bibitem[\protect\citeauthoryear{Evans et al.}{2009}]{evans2009}
Evans, N. J., Dunham, M. M., J{\o}rgensen, J. K., et al. 2009, ApJS, 181, 321-
350

\bibitem[\protect\citeauthoryear{Feiden}{2016}]{feiden16}
Feiden, G. A. 2016, A\&A, 593, A99


\bibitem[\protect\citeauthoryear{Frasca et al.}{2015}]{frasca15}
Frasca, A., Biazzo, K., Lanzafame, A. C., et al. 2015, \aap, 575, A4

\bibitem[\protect\citeauthoryear{Frasca et al.}{2017}]{frasca17}
Frasca, A., Biazzo, K., Alcal\'a, J. M., et al. 2017, \aap, 602, A33

\bibitem[\protect\citeauthoryear{\textit{Gaia} Collaboration}{2018}]{gaia}
\textit{Gaia} Collaboration, Brown, A. G. A., Vallenari, A., et al. 2018, A\&A, 616, A1

\bibitem[\protect\citeauthoryear{\textit{Gaia} Collaboration}{2021}]{gaiaedr3}
\textit{Gaia} Collaboration, Brown, A. G. A., Vallenari, A., et al. 2021, A\&A, 649, A1

\bibitem[\protect\citeauthoryear{Gagn\'e et al.}{2018}]{gagne18}
Gagn\'e, J., Mamajek, E. E., Malo, L., et al. 2018a, ApJ, 856, 23

\bibitem[\protect\citeauthoryear{Gangi et al.}{2022}]{gangi22}
Gangi, M., Antoniucci, S., Biazzo, K. et al. 2022, in press (arXiv:2208.14895)

\bibitem[\protect\citeauthoryear{Galli et al.}{2013}]{galli13}
Galli, P. A. B., Bertout, C., Teixeira, R., Ducourant, C. 2013, A\&A, 558, A77

\bibitem[\protect\citeauthoryear{Galli et al.}{2020}]{galli2020}
Galli, P. A. B., Bouy, H., Olivares, J., et al. 2020, A\&A 643, A148


\bibitem[\protect\citeauthoryear{Gullikson et al.}{2014}]{gullikson14}
Gullikson, K., Dodson-Robinson, S., \& Kraus, A. 2014, AJ, 148, 53


\bibitem[\protect\citeauthoryear{Herczeg \& Hillenbrand}{2014}]{hh14}
Herczeg, G. J., \& Hillenbrand, L. A. 2014, ApJ, 786, 97

\bibitem[\protect\citeauthoryear{Hughes et al.}{1994}]{hughes1994}
Hughes, S.M. G., Gear, W. K., \& Robson, E. I. 1994, ApJ, 428, 143


\bibitem[\protect\citeauthoryear{Krautter et al.}{1997}]{krautter1997}
Krautter, J., Wichmann, R., Schmit, J. H. M. M., et al. 1997, A\&ASS, 123,
329

\bibitem[\protect\citeauthoryear{Lazzoni et al.}{2020}]{lazzoni2020}
Lazzoni, C., Gratton, R., Alcal\'a, J., et al. 2020, A\&A, 635, L11 


\bibitem[\protect\citeauthoryear{Majidi et al.}{2020}]{majidi2020}
Majidi, F. Z., Desidera, S., Alcal\'a, J. M. et al. et al. 2020, A\&A, 644, A169

\bibitem[\protect\citeauthoryear{Manara et al.}{2022}]{manara2022}
Manara, C.F., Ansdell, M., Rosotti, G.P., et al. 2022, arXiv:2203.09930 [astro-ph.SR]

\bibitem[\protect\citeauthoryear{Manara et al.}{2013}]{manara13}
Manara, C. F., Testi, L., Rigliaco, E., et al. 2013, A\&A, 551, A107


\bibitem[\protect\citeauthoryear{Manara et al.}{2017b}]{manara17b}
Manara, C. F., Frasca, A., Alcalá, J. M., et al. 2017b, A\&A, 605, A86


\bibitem[\protect\citeauthoryear{Manara et al.}{2018}]{manara18}
Manara, C. F., Prusti, T., Comeron, F., et al. 2018, A\&A, 615, L1



\bibitem[\protect\citeauthoryear{Mer\'in et al.}{2008}]{merin2008}
Mer\'in, B., J{\o}rgensen, J., Spezzi, L., et al. 2008, ApJS, 177, 551

\bibitem[\protect\citeauthoryear{Mortier et al.}{2011}]{mortier11}
Mortier, A., Oliveira, I., \& van Dishoeck, E. F. 2011, MNRAS, 418, 1194

\bibitem[\protect\citeauthoryear{Nardiello et al.}{2020}]{mimmo20}
Nardiello, D., Piotto, G., Deleuil, M., et al. 2020, MNRAS, 495, 4924

\bibitem[\protect\citeauthoryear{Palla et al.}{2007}]{pallaetal2007} 
Palla, F., Randich, S., Pavlenko, Ya. V., Flaccomio, E., Pallavicini, R. 2007, \apj, 659, 41L

\bibitem[\protect\citeauthoryear{Pastorelli et al.}{2019}]{pastorelli19}
Pastorelli, G., Marigo, P., Girardi, L., et al. 2019, MNRAS, 485, 5666

\bibitem[\protect\citeauthoryear{Pastorelli et al.}{2020}]{pastorelli20}
Pastorelli, G., Marigo, P., Girardi, L., et al. 2020, MNRAS, 498, 3283

\bibitem[\protect\citeauthoryear{Paxton et al.}{2015}]{paxton15}
Paxton, B., Marchant, P., Schwab, J., et al. 2015, ApJS, 220, 15

\bibitem[\protect\citeauthoryear{Pecaut \& Mamajek}{2013}]{pm13}
Pecaut, M. J., Mamajek, E. E. 2013, ApJS, 208, 9

\bibitem[\protect\citeauthoryear{Pecaut \& Mamajek}{2016}]{pm16}
Pecaut, M. J., Mamajek, E. E. 2016, MNRAS, 461, 794

\bibitem[\protect\citeauthoryear{Prisinzano et al.}{2022}]{prisinzano22}
Prisinzano, L., Damiani, F., Sciortino, S. et al. 2022, \aa, 664, 175


\bibitem[\protect\citeauthoryear{Riddick et al.}{2007}]{riddick2007}
Riddick, F., Roche, P., \& Lucas, P. 2007, MNRAS, 381, 1067


\bibitem[\protect\citeauthoryear{Rygl et al.}{2012}]{rygl2012}
Rygl, K. L. J., Brunthaler, A., Sanna, A., et al. 2012, A\&A, 539, A79

\bibitem[\protect\citeauthoryear{Sacco et al.}{2007}]{saccoetal2007} 
Sacco, G. G., Randich, S., Franciosini, E., Pallavicini, R., \& Palla, F. 2007, \aap, 462, L23

\bibitem[\protect\citeauthoryear{Sicilia-Aguilar et al.}{2014}]{s-a14}
Sicilia-Aguilar, A., Roccatagliata, V., Getman, K., et al. 2014, A\&A, 562, A131

\bibitem[\protect\citeauthoryear{Smette et al.}{2015}]{smette15}
Smette, A., Sana, H., Noll, S., et al. 2015, A\&A, 


\bibitem[\protect\citeauthoryear{Samatellos \& Herczeg}{2015}]{staher15}
Stamatellos, D., Herczeg, G. J. 2015, \mnras,  449, 3432

\bibitem[\protect\citeauthoryear{Spezzi et al.}{2011}]{spezzi11}
Spezzi, L., Vernazza, P., Mer\'in, B., et al. 2011, ApJ, 730, 65

\bibitem[\protect\citeauthoryear{Tody}{1986}]{iraf1}
Tody, D. 1986, SPIE Conf. Ser., 627, 733

\bibitem[\protect\citeauthoryear{Tody}{1993}]{iraf2}
Tody, D. 1993, ASP Conf. Ser., 52, 173

\bibitem[\protect\citeauthoryear{Torres et al.}{2006}]{torres2006}
Torres, C. A. O., Quast, G. R., da Silva, L., et al. 2006, A\&A, 460, 695--708  

\bibitem[\protect\citeauthoryear{Vernet et al.}{2011}]{vernet11}
Vernet, J., Dekker, H., D'Odorico, S., et al., 2011, \aap, 536, A105 

\bibitem[\protect\citeauthoryear{Vorobyov \& Basu}{2009}]{vb2009}
Vorobyov, E. I., \& Basu, S. 2009, ApJ, 703, 922


\bibitem[\protect\citeauthoryear{White \& Basri}{2003}]{white2003}
White, R. J., Basri, G. 2003, ApJ, 582, 1109

\bibitem[\protect\citeauthoryear{Zari et al.}{2018}]{zari18}
Zari, E., Hashemi, H., Brown, A. G. A., et al. 2018, A\&A, 620, A172

\bibitem[\protect\citeauthoryear{Zhou et al.}{2014}]{zhou2014}
Zhou, Y., Herczeg, G. J., Kraus, A. L. et al. 2014, \apj, 783, 17

\end{thebibliography}
\end{document}